\documentclass[%
 showkeys,
 reprint,
superscriptaddress,
 amsmath,amssymb,
 aps,
]{revtex4-2}

\usepackage{graphicx}
\usepackage{dcolumn}
\usepackage{bm}

\usepackage[per-mode=symbol]{siunitx}%
\usepackage{nicefrac}
\usepackage[colorlinks=true, linkcolor=blue, citecolor=blue, urlcolor=blue]{hyperref}

\newcommand{\CuOSeO}{$\text{Cu}_2 \text{OSeO}_3$}

\newcommand{\dBdz}{\text{d}B_{z}/\text{d}z}

\DeclareSIUnit\Phizero{\text{\ensuremath{\Phi_0}}}

\def\ekutaffil{%
  \affiliation{%
    Physikalisches Institut, Center for Quantum Science (CQ) and LISA$^+$,
    University of T\"ubingen,
    72076 T\"ubingen,
    Germany
    }
}

\def\ubasaffil{%
  \affiliation{%
    Department of Physics,
    University of Basel,
    4056 Basel,
    Switzerland
    }
}

\def\snsiaffil{%
  \affiliation{%
    Swiss Nanoscience Institute,
    University of Basel,
    4056 Basel,
    Switzerland
    }
}

\def\ibmaffil{%
  \affiliation{%
    IBM Research Europe – Z\"urich,
    8803 R\"uschlikon,
    Switzerland
    }
}

\def\ptbaffil{%
  \affiliation{%
    Department Quantum Electronics,
    Physikalisch-Technische Bundesanstalt (PTB),
    38116 Braunschweig,
    Germany
    }
}

\def\epflaffil{%
  \affiliation{%
    Institute of Physics,
    École Polytechnique Fédérale de Lausanne,
    1015 Lausanne,
    Switzerland
    }
}

\begin{document}


\title{Advanced SQUID-on-lever scanning probe for high-sensitivity magnetic microscopy with sub-100-nm spatial resolution}


\author{Timur Weber} 
\altaffiliation[]{These authors contributed equally to this work.}
\ekutaffil

\author{Daniel Jetter} 
\altaffiliation[]{These authors contributed equally to this work.}
\ubasaffil

\author{Jan Ullmann} 
\ekutaffil
\author{Simon A. Koch} 
\ekutaffil
\author{Simon F. Pfander} 
\ekutaffil

\author{Katharina Kress} 
\ubasaffil
\author{Andriani Vervelaki} 
\ubasaffil
\author{Boris Gross} 
\ubasaffil

\author{Oliver Kieler} 
\ptbaffil

\author{Ute Drechsler} 
\ibmaffil

\author{Priya R. Baral} 
\epflaffil
\author{Arnaud Magrez} 
\epflaffil

\author{Reinhold Kleiner} 
\ekutaffil

\author{Armin W. Knoll} 
\ibmaffil

\author{Martino Poggio} 
\ubasaffil
\snsiaffil

\author{Dieter Koelle}
\email[Contact author: ]{koelle@uni-tuebingen.de}
\ekutaffil


\begin{abstract}
Superconducting quantum interference devices (SQUIDs) are exceptionally sensitive magnetometers capable of detecting weak magnetic fields. Miniaturizing these devices and integrating them onto scanning probes enables high-resolution imaging at low-temperature. Here, we fabricate nanometer-scale niobium SQUIDs with inner-loop sizes down to $\SI{10}{\nano\meter}$ at the apex of individual planar silicon cantilevers via a combination of wafer-scale optical lithography and focused-ion-beam (FIB) milling. These robust SQUID-on-lever probes overcome many of the limitations of existing devices, achieving spatial resolution better than 100~nm, magnetic flux sensitivity of \SI{0.3}{\micro\Phizero\per\sqrt\hertz}, and operation in magnetic fields up to about 0.5~T at 4.2~K. Nanopatterning via Ne- or He-FIB allows for the incorporation of a modulation line for coupling magnetic flux into the SQUID or a third Josephson junction for shifting its phase. Such advanced functionality, combined with high spatial resolution, large magnetic field range, and the ease of use of a cantilever-based scanning probe, extends the applicability of scanning SQUID microscopy to a wide range of magnetic, normal conducting, superconducting, and quantum Hall systems. We demonstrate  magnetic imaging of skyrmions at the surface of bulk \CuOSeO. Analysis of the point spread function determined from imaging a single skyrmion yields a full-width-half-maximum of 87\,nm. Moreover, we image modulated magnetization patterns with a period of 65\,nm.
%
%
\end{abstract}

\keywords{superconductivity, SQUID, scanning probe microscopy, magnetic imaging}


\maketitle


\section{Introduction}\label{sec1} 

Magnetic microscopy -- because of its ability to reveal magnetization patterns and current distributions -- is crucial for studying the properties of magnetic, normal conducting, and superconducting materials. Unlike bulk measurements, which integrate over entire samples, magnetic imaging reveals information about length scales, inhomogeneities, and interactions. A wide variety of techniques, including scanning probe microscopy, light microscopy, and electron microscopy, are available for an even wider variety of materials and conditions~\cite{christensen_2024_2024}. Scanning superconducting quantum interference device (SQUID) microscopy (SSM) is particularly well-suited for sensitive and high-resolution imaging of transport and magnetization at low temperatures. 

In recent years, SSM has been used to explore magnetism in nanostructures~\cite{vasyukov_imaging_2018,wyss_stray-field_2019}, at complex oxide interfaces~\cite{bert_direct_2011,christensen_strain-tunable_2019,wang_imaging_2015}, in topological insulators~\cite{lachman_visualization_2015}, and in few-layer 2-dimensional (2D) magnets~\cite{noah_nano-patterned_2023,zur_magnetic_2023,vervelaki_visualizing_2024,bagani_imaging_2024}, as well as allowing for the imaging of orbital magnetism in bilayer graphene~\cite{tschirhart_imaging_2021}. SSM was also used to image current flow in both a quantum spin Hall~\cite{nowack_imaging_2013} and a quantum anomalous Hall insulator~\cite{ferguson_direct_2023}, in graphene in the quantum Hall regime~\cite{uri_nanoscale_2020}, in magic-angle twisted bilayer graphene~\cite{uri_mapping_2020}, at oxide domain walls~\cite{kalisky_locally_2013}, and in WTe$_2$ in a regime where electrons flow hydrodynamically~\cite{aharon-steinberg_direct_2022}. In the area of superconductivity, SSM has long been a central tool for discovery, e.g.~on the nature of the order parameter symmetry in cuprate superconductors~\cite{persky_studying_2022,kirtley_fundamental_2010}. Most recently, SSM was used to observe the flow of superconducting vortices in a Pb film~\cite{embon_imaging_2017} and quantum fluctuations near criticality in NbTiN~\cite{kremen_imaging_2018}. It enabled the discovery of a hidden magnetic phase above the superconducting transition temperature of the van der Waals superconductor 4Hb-TaS$_2$~\cite{persky_magnetic_2022}, as well as vortex excitations carrying a temperature-dependent fraction of the flux quantum $\Phi_0$~\cite{iguchi_superconducting_2023}.

At the moment, there are two main types of high-resolution SSM. The first is based on a so-called SQUID-on-chip sensor, in which the superconducting device is fabricated via standard lithography on a chip, and a small, spatially separated pick-up loop, often at the corner of the chip, couples the sensor to the sample of interest~\cite{kirtley_highresolution_1995,koshnick_terraced_2008,kirtley_scanning_2016}. These SQUIDs can be fabricated in large quantities and have the advantage of including on-chip circuitry such as flux modulation coils and a coil for magnetic susceptibility measurements. However, their spatial resolution is typically in the micrometer range, due to the size of the pick-up loop and limitations on the minimum SQUID-to-sample distance~\cite{persky_studying_2022}; moreover, they only function in magnetic fields up to about 10~mT, because of the properties of the Nb/AlO$_x$/Nb Josephson junctions (JJs)~\cite{huber_gradiometric_2008}. A second type of sensor, known as a SQUID-on-tip, consists of a SQUID loop that is fabricated directly at the apex of a quartz capillary via a three-step deposition of superconducting material~\cite{finkler_self-aligned_2010,vasyukov_scanning_2013}. The JJs of these SQUIDs are Dayem bridges (or constriction junctions), allowing for SQUIDs with diameters less than 100~nm and critical fields in the Tesla range~\cite{bagani_sputtered_2019}. This combination results in SSM probes with sub-micrometer spatial resolution that operate in a high magnetic field beyond \SI{1}{\tesla}. However, the process by which SQUID-on-tips are fabricated precludes the application of conventional wafer-scale processing and the integration of circuitry such as a modulation coil for controlling the flux coupling into the SQUID or for local magnetic susceptibility measurements. Moreover, the choice of superconducting materials is limited by the non-planar surfaces on which they must be deposited or grown. Magnetic flux sensitivities of both types of SSM probes typically range from hundreds of n$\Phi_0/\sqrt{\text{Hz}}$ to a few \textmu$\Phi_0/\sqrt{\text{Hz}}$ at 4.2~K~\cite{jose_martinez-perez_nanosquids_2017}.

Recently, nanometer-scale SQUIDs based on Dayem bridge JJs have been realized at the tip of modified commercial Si cantilevers for atomic force microscopy (AFM) using a process based on focused ion beam milling (FIB)~\cite{wyss_magnetic_2022}. Here, we go significantly beyond this initial proof-of-concept by realizing Nb nanoSQUIDs that are integrated on custom-made Si cantilevers. We use wafer-scale thin-film deposition and optical lithography to produce cantilever chips that contain prepatterned Nb microstructures on planar cantilevers. We combine this with a final FIB milling step, to fabricate sensors reproducibly and in large quantities. This process yields robust SQUID-on-lever sensors that allow for a well-controlled surface approach, magnetic imaging with spatial resolution better than 100~nm, a magnetic flux sensitivity of 0.3~\textmu$\Phi_0/\sqrt{\text{Hz}}$, operation in magnetic fields up to $\sim 0.5$~T, and the integration of on-tip circuitry, including a modulation line or a third JJ to flux-bias or phase-bias the devices at their optimum working point. We demonstrate the capabilities of these sensors by imaging nanometer-scale magnetic configurations at the surface of the chiral magnet \CuOSeO. Endowed with the advantages of both SQUID-on-chip and SQUID-on-tip sensors, the resulting probes promise to expand the reach of SSM to a variety of nanometer-scale magnetic, normal conducting, and superconducting systems.

\section{Sensor fabrication and layout} \label{sec2} 

\begin{figure*}[t]
\centering
\includegraphics[width=1\textwidth]{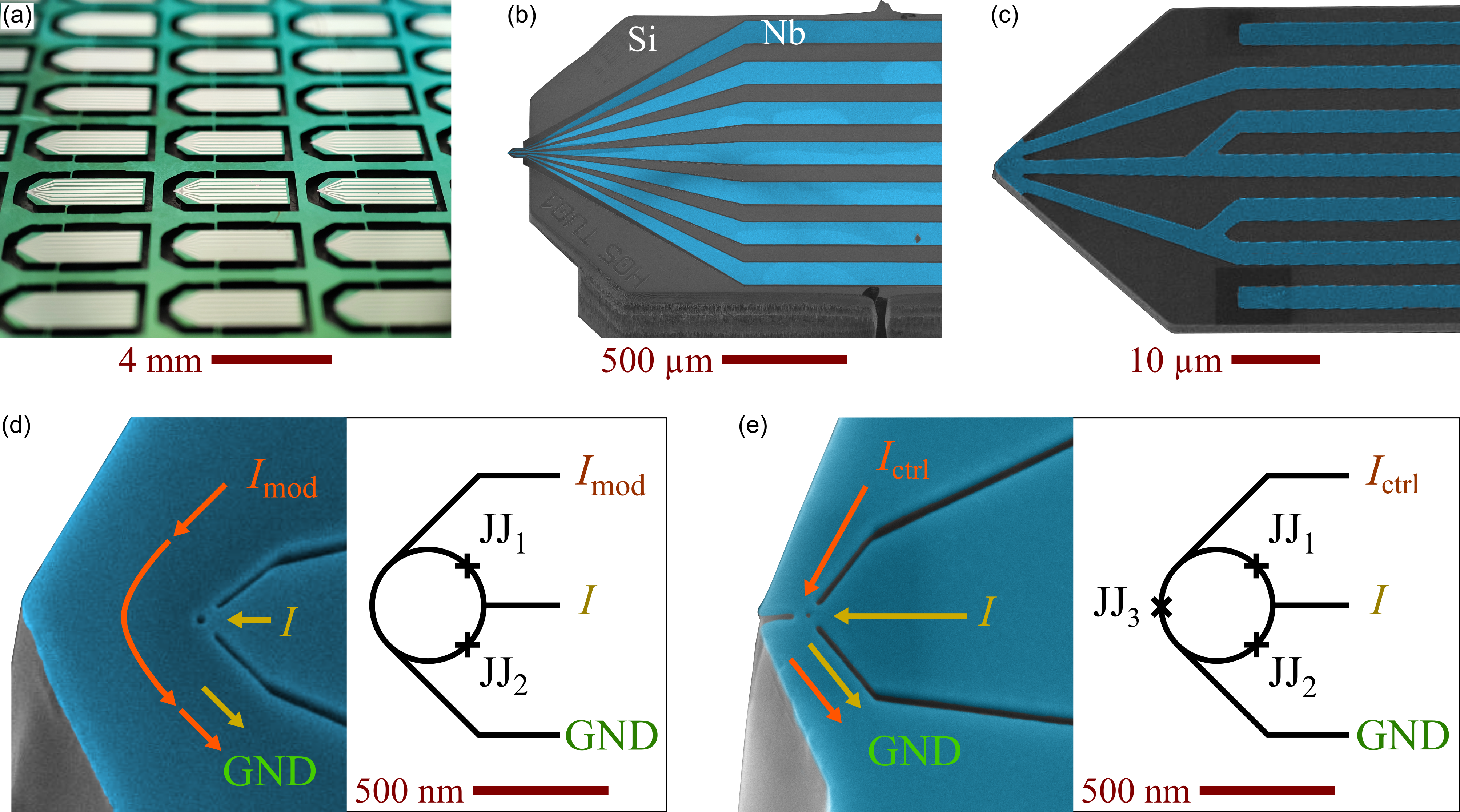}
\caption{Planar SQUID-on-lever with advanced functionality.
(a) Optical micrograph of part of an SOI wafer with 21 patterned cantilever chips, which can be individually broken out. 
(b) False-colored scanning electron microscopy (SEM) image of part of a cantilever chip with patterned Nb lines extending to the cantilever (left) that protrudes beyond the chip body. 
(c) False-colored SEM image of a cantilever. Three Nb lines (blue) connect to a Nb triangle at the apex (left). 
(d), (e): False-colored SEM images (left) and corresponding circuit diagrams (right) illustrate examples of SQUID circuits that were subsequently patterned via Ne- and He-FIB milling by cutting trenches and a hole (diameter $d=36$~nm in (d) and 19~nm in (e)) into the Nb triangle. In (d), a modulation current \(I_\mathrm{mod}\) can be applied to couple flux into the 2-JJ SQUID. In (e) a control current \(I_\mathrm{ctrl}\) can be used to shift the phase of the 3-JJ SQUID. Each SEM image has been taken at an angle of 45\,$^\circ$.}
\label{fig1}
\end{figure*}

We fabricate the probes shown in Fig.~\ref{fig1} on 4-inch Silicon-on-insulator (SOI) wafers resulting in approximately 500 cantilever chips per wafer. Each chip has a size of nominally 4~mm~$\times$~1.5~mm (see Fig.~\ref{fig1}(a)). This wafer-scale process starts with a thermal oxidation step, which results in a 50-nm-thick silicon-oxide at the surface. Next, we sputter 50~nm of Nb, capped with a 10-nm-thick protective layer of alumina (Al$_2$O$_3$). We then pattern the Nb film, cantilever, and chip body via optical lithography and reactive ion etch (RIE). To define the Nb microstructures, we use Cl$_2$/BCl$_3$/Ar, Cl$_2$/Ar, and Ar/CHF$_3$ RIE to etch the alumina, Nb, and thermal oxide, respectively. We use SF$_6$/C$_4$F$_8$ deep RIE to etch both the cantilever shape out of the Si device layer and the chip body from the backside using the buried oxide of the SOI wafer as an etch-stop. Finally, we release the cantilevers by removing the buried oxide locally using buffered HF. The resulting cantilevers are about \SI{60}{\micro\meter} long, \SI{40}{\micro\meter} wide, and \SI{2}{\micro\meter} thick with spring constant of $\sim$~\SI{45}{\newton\per\meter}. As shown in Fig.~\ref{fig1}(b),(c), the cantilevers include patterned Nb leads connecting to a triangular device area at the apex of the probe.

At the apex of each cantilever, we then pattern the final sensing circuit using high-resolution Ne- and He-FIB milling. 
First, we use Ne-FIB to shape the cantilever tip for optimal AFM operation. If required, we can also remove Nb around the sensing area and the edges of the film.
This removes material damaged by the etching process, further limits flux-focusing effects and prevents the entrance of superconducting vortices close to the SQUID during operation in an applied magnetic field. 
Once the tip and sensor area have been milled to the appropriate geometry, a specific pattern is applied to fabricate the intended SQUID circuit. By milling away Nb, we define physical gaps in the superconducting film, allowing us to pattern superconducting leads, loops, and Dayem bridge JJs. As part of the patterning procedure, we measure and correct for thermal and mechanical drift of the ion beam, to avoid blurring of the desired features.

From studies of the dependence of the critical current $I_\text{c}$ on the width $w$ of single JJs, we find that we can realize JJs with $w$ down to $\sim 30~$nm via Ne-FIB milling, while with He-FIB we can produce JJs with even smaller $w$ -- down to $\sim 10~$nm -- before $I_\text{c}$ drops to zero at 4.2~K~\cite{weber_niobium_2025}. It seems that He-FIB milling produces less edge damage than Ne-FIB milling, i.e, He-FIB allows us to realize smaller holes and narrower JJs with reasonable $I_\text{c}$, at the cost of lower milling rates. This process results in geometric features as small as 10~nm.

The triangular Nb structure at the apex of the cantilever is connected by three Nb lines (see  Fig.~\ref{fig1}(c)), that are defined by two lithographically patterned trenches between them.
To pattern a SQUID with two JJs (JJ$_1$ and JJ$_2$ -- see Fig.~\ref{fig1}(d)), we extend those two trenches by FIB milling towards the cantilever tip. We then mill a SQUID hole with diameter $d$ between them. The distance between the SQUID hole and each trench defines the widths $w_1$ and $w_2$ of JJ$_1$ and JJ$_2$, respectively. The Ne-FIB-cut trenches are 50~nm wide up to a few hundred nanometers from the SQUID hole, where they are reduced to a width of 25~nm. The final width of the trenches near the SQUID hole defines the length of the Dayem bridge JJs. Using He-FIB, the width of the trenches can be reduced to 15~nm for the last 50 to 70~nm. 

For readout, a bias current $I$ flows from the central Nb line through the JJs to ground, shown in Fig.~\ref{fig1}(d) as the bottom Nb line. In addition, a modulation current $I_\mathrm{mod}$ can be sent from the upper Nb line -- flowing along part of the SQUID loop -- to ground. When the SQUID is operated in the finite voltage state, the modulation current does not flow across the junctions. Via $I_\mathrm{mod}$, the SQUID can be flux-biased and operated in a flux-locked loop (FLL).

With a small modification, i.e.\ by adding a third FIB cut, which runs from the apex towards the SQUID hole, the layout described above can be converted into a SQUID with three JJs (see Fig.~\ref{fig1}(e))~\cite{meltzer_multi-terminal_2016, uri_electrically_2016,wolter_static_2022}. In that case, we denote the current that flows from the upper Nb line to ground as the control current $I_\mathrm{ctrl}$. In general, both $I$ and $I_\mathrm{ctrl}$ can flow across all three JJs. For fixed $I$, the control current shifts the phase differences $\delta_i$ of the superconducting order parameter across the three junctions JJ$_i$ in the SQUID loop. The control current thus provides a phase bias that can be used to shift the dependence of the critical current $I_c$ on the applied flux $\Phi$ or of the voltage $V$ on $\Phi$. This feature can be implemented in a phase-locked loop (PLL) to keep the SQUID at its optimum working point.

\section{Electrical characterization} \label{sec3} 

\begin{figure*}[t]%
\centering
\includegraphics[width=1\textwidth]{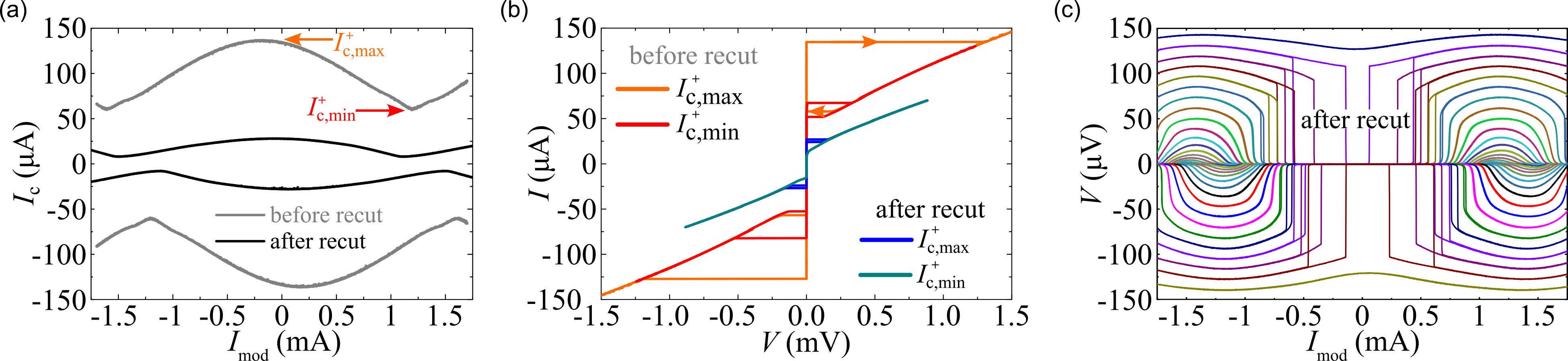}
\caption{Electric transport properties of 2-JJ SQUID-on-lever.
4-point measurements at $T=4.2$~K for a SQUID milled by Ne-FIB 
with hole diameter $d = 36$~nm and JJ widths $w_1 = w_2 = 55$~nm before and  $w_1 = 48$~nm and $w_2 = 42$~nm after recutting by Ne-FIB.
(a) Critical current $I_\text{c}$ vs modulation current $I_\text{mod}$ before and after recutting the JJs. Arrows indicate maximum and minimum positive critical currents $I^+_\text{c,max}$ and $I^+_\text{c,min}$, respectively.  
(b) Current-voltage characteristics before and after recutting, recorded at $I_\text{mod}$ values that provide $I^+_\text{c,max}$ and $I^+_\text{c,min}$.
Before recutting, the IVCs show a pronounced hysteresis (arrows indicate sweep direction of current), which is almost absent after recutting.
%
(c)
Voltage $V$ vs $I_\text{mod}$ oscillations, measured after recutting for different bias currents (from $-25$ to 25~\textmu A in 1~\textmu A steps).}
\label{fig2}
\end{figure*}

The electrical properties and magnetic field response of the SQUID-on-lever (SoL) sensors are characterized at temperature $T=4.2~$K. A magnetic field $B_\text{a}=\mu_0H_\text{a}$ is applied perpendicular to the thin film plane. Below, we present and discuss results obtained on two representative sensors: the first 
is a SoL with two JJs (2JJ-SoL), which includes a modulation line and was patterned via Ne-FIB;
the second 
is a SoL with three JJs (3JJ-SoL), patterned via both Ne- and He-FIB to obtain a very small SQUID hole diameter.

Figure \ref{fig2} summarizes results from 4-point measurements of current $I$ vs voltage $V$ characteristics (IVCs) of the 2JJ-SoL 
with a hole diameter $d=36$~nm. The two JJs originally had a width $w_{1,2}=55~$nm; in a second Ne-FIB milling step (recut) we reduced the JJs widths down to $w_1=48~$nm and $w_2=42~$nm, to reduce hysteresis in their IVCs. We note that the quoted JJ widths are nominal widths based on SEM images, which do not include the reduction of the effective JJ width due to edge damage induced by Ne-FIB milling. Based on the scaling of experimentally determined critical currents with geometric JJ width $w$ for our 2JJ-SoLs cut by Ne-FIB milling, we roughly estimate that the degradation of superconductivity extends laterally from the milled edges into the superconductor by a damage depth of around $15-20$~nm; this effectively reduces the JJ widths by around $30-40$~nm. 

Figure \ref{fig2}(a) shows the oscillation of the critical current $I_\text{c}$ vs modulation current $I_\text{mod}$ before (grey curves) and after (black curves) recut for both current polarities. In the following, we analyze those patterns, in particular with respect to the determination of the dimensionless screening parameter $\beta_L\equiv 2LI_0/\Phi_0$ (with SQUID inductance $L$, magnetic flux quantum $\Phi_0\approx 2.068\times 10^{-15}~$Vs and noise-free critical current $I_0\approx I_\text{c,max}/2$ per JJ)~\cite{tesche_dc_1977, chesca_SQUID_2004} 

Before recut, the positive branches of the $I_\text{c}(I_\text{mod})$ curves yield maximum and minimum critical currents $I^+_\text{c,max}=135~$\textmu A (at $I_\text{mod}=-185~$\textmu A) and $I^+_\text{c,min}=67~$\textmu A (at $I_\text{mod}=1.2~$mA), respectively; this corresponds to a normalized modulation depth $\Delta i^+_\text{c}=(I^+_\text{c,max}-I^+_\text{c,min})/I^+_\text{c,max}=0.50$. As a first order approximation, numerical simulations for a symmetric SQUID with a sinusoidal current-phase relation (CPR) of the JJs and negligible thermal noise, predict that $\Delta i^+_\text{c}=0.5$ would correspond to $\beta_L=1$~\cite{tesche_dc_1977, chesca_SQUID_2004}. The oscillation period $\Delta I_\text{mod}=2.8~$mA of the $I_\text{c}(I_\text{mod})$ curves corresponds to a flux change of $\Phi_0$ in the SQUID. Accordingly, the mutual inductance of the modulation line is given by $M=\Phi_0/\Delta I_\text{mod}=0.74~$pH. The relative shift $\Delta\Phi/\Phi_0\approx 0.13$ of the $I_\text{c}$ oscillations for both polarities along the flux axis can be attributed to a critical current asymmetry $\alpha_I$ in the two JJs with critical currents $I_\text{c,i}=I_0(1\pm\alpha_I)$ ($i=1,2$), with $\Delta\Phi/\Phi_0=\alpha_I\beta_L$ (for negligible  inductance asymmetry)~\cite{chesca_SQUID_2004}. Both, an $I_\text{c}$ asymmetry and a deviation from a sinusoidal CPR reduce the modulation depth of $I_\text{c}(\Phi)$~\cite{tesche_dc_1977,prance_sensitivity_2023}, which means that the above estimate of $\beta_L=1$ (and hence the estimate of $L$) is likely too large. This indeed seems to be the case, as the measured $I_\text{c,max}=135~$\textmu A and $\beta_L=1$ yields $L=15~$pH, whereas our numerical simulations of the circulating supercurrent distribution~\cite{khapaev_current_2002} for the given SQUID geometry (with London penetration depth $\lambda_\text{L}=110~$nm) predict $L=1.1~$pH without edge damage and $L=2.8\,(2.1)~$pH with 20 (15)~nm damage depth. Those much lower values for $L$ are consistent with the measured $M=0.74~$pH, which is expected to be a significant fraction of $L$ and result in an estimated $\beta_L\approx 0.07$ without edge damage and 0.18 (0.13) with 20 (15)~nm damage depth.

\begin{figure*}[t]%
\centering
\includegraphics[width=1\textwidth]{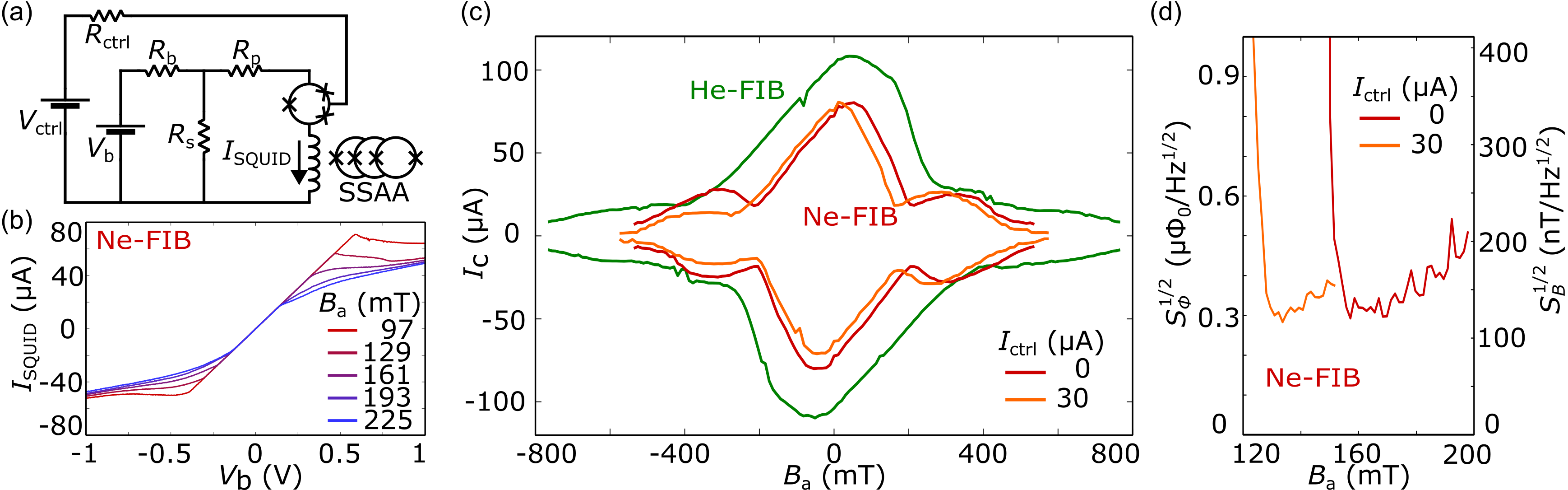}
\caption{Electric transport and noise properties of 3JJ-SoL.
Results obtained at $T=\SI{4.2}{K}$ with the configuration in (a) are shown in (b)--(d) for a 3JJ-SoL milled by Ne-FIB 
($w_1 = w_2 = \SI{50}{nm}$, $w_3 = \SI{75}{nm}$); (c) also includes results from a 3JJ-SoL milled by He-FIB 
($w_1 = w_2 = \SI{28}{nm}$, $w_3 = \SI{37}{nm}$), which was later used for imaging. Both devices have a hole diameter $d = \SI{15}{nm}$.
(a) Readout circuit used for 3JJ-SoL characterization and imaging. 
(b) $I_\text{SQUID}(V_\text{b})$ curves at various magnetic fields $B$, with $I_\text{ctrl}=0$ for the 3JJ-SoL milled by Ne-FIB ($R_\text{b}=6.2~\text{k}\Omega$, $R_\text{s}=1~\Omega$).
(c) $I_\text{c}(B)$ for the 3JJ-SoL milled by He-FIB with $I_\text{ctrl}=0$ and for the 3JJ-SoL milled by Ne-FIB with $I_\text{ctrl}=0$ and 30~\textmu A.
(d) Rms spectral densities of flux noise $S_\Phi^{\nicefrac{1}{2}}$ and field noise $S_B^{\nicefrac{1}{2}}$ at 12~kHz vs applied magnetic field $B$ for the 3JJ-SoL milled by Ne-FIB. The two curves correspond to the most sensitive working points found with $I_\text{ctrl}=0$ at $V_\text{b}=0.15~$V and with $I_\text{ctrl}=30~$\textmu A at $V_\text{b}=0.4~$V.}
\label{fig3}
\end{figure*}

After recut, the critical currents are significantly reduced to $I^+_\text{c,max}=27~$\textmu A (at $I_\text{mod}=-60~$\textmu A) and $I^+_\text{c,min}=8~$\textmu A (at $I_\text{mod}=1.1~$mA). This yields a normalized modulation depth $\Delta i^+_\text{c}=0.70$; for a symmetric SQUID with sinusoidal CPR of the JJs and negligible thermal noise, this would correspond to $\beta_L=0.42$. From the slightly reduced oscillation period $\Delta I_\text{mod}=2.6~$mA of the $I_\text{c}(I_\text{mod})$ curves, we extract a slightly increased $M=0.8~$pH. A relative shift $\Delta\Phi/\Phi_0\approx 0.05$ of the $I_\text{c}$ oscillations for both polarities indicates a finite critical current asymmetry. Again, we expect that $\beta_L$ is much smaller than 0.42 due to $I_\text{c}$ asymmetry and deviation from a sinusoidal CPR; in fact, numerical simulations (again with $\lambda_\text{L}=110~$nm) predict $L=1.2~$pH without edge damage and $L=5.4\,(2.6)~$pH with 20 (15)~nm damage depth. With the measured $I^+_\text{c,max}=27~$\textmu A, those estimated inductance values result in $\beta_L\approx 0.02$ without edge damage and 0.07 (0.03) with 20 (15)~nm damage depth.

Figure \ref{fig2}(b) shows IVCs of the 2JJ-SoL recorded for both current-sweep directions before and after recut, at values of $I_\text{mod}$ that were adjusted to yield the critical currents $I^+_\text{c,max}$ and $I^+_\text{c,min}$. 
Before recut, both IVCs show a significant hysteresis which we attribute to Joule heating effects.
To reduce heating effects, we have recut the JJs to reduce their widths and accordingly their critical currents. As is evident from Fig.~\ref{fig2}(b), the trimming of the JJs strongly suppresses the hysteresis in the IVCs. This ability to controllably cut and recut features, allows for a fine tunability of the SQUID-on-lever properties and demonstrates the robustness of the fabrication scheme.

Figure \ref{fig2}(c) shows a family of voltage $V$ vs~modulation current $I_\text{mod}$ curves for different bias currents $I$, measured after recutting the 2JJ-SoL. We cannot find a shift of the voltage maxima along the $I_\text{mod}$ axis with increasing $I$; this indicates a negligible asymmetry in the SQUID inductance $L$. For $|I_\text{mod}|\lesssim 0.5~$mA, the hysteresis in the IVCs causes the vertical switching steps in $V(I_\text{mod})$, which prevents operation of the SQUID in this regime. However, outside  this regime ($|I_\text{mod}|\gtrsim 0.5~$mA), the SQUID can be flux-biased via $I_\text{mod}$ in a stable working point, and it can be operated as a flux-to-voltage transducer in a simple direct voltage readout scheme. For example, for $I=19~$\textmu A and $I_\text{mod}=0.72~$mA, a very high transfer function $\frac{\partial V}{\partial\Phi}\equiv V_\Phi = 9~\frac{\text{mV}}{\Phi_0}$ is reached. The integration of a modulation line in the SoL layout is an important feature, as it allows us to operate the sensor in its optimum working point and to use conventional SQUID readout electronics with FLL operation to maintain the optimum working point, even in variable external magnetic field. We note, however, that $I_\text{mod}$ in the mA range could be too invasive for some applications, as it can produce a disturbing magnetic field at the surface of a sample that is imaged. For example, a straight wire carrying 1~mA produces a Biot-Savart field of 4~mT at a radial distance of 50~nm. A solution of this problem can be provided by the 3JJ-SoLs.  

Figure \ref{fig3} summarizes results from measurements of two 3JJ-SoLs patterned with either Ne- or He-FIB, both with hole diameter $d=15~$nm. For those devices, we did not trim the JJ widths to avoid hysteresis in the IVCs. Instead, we operated them using the readout scheme, shown in Fig.~\ref{fig3}(a), which is implemented in the scanning SQUID microscope. This scheme allows for readout of a SQUID even if its IVC (with current bias) is hysteretic, which therefore enables SQUID operation over the widest possible magnetic field range~\cite{vasyukov_scanning_2013,wyss_magnetic_2022}. This circuit operates the SQUIDs in voltage-bias mode at bias voltage $V_\text{b}$ with a large bias resistor $R_\text{b}$, a small shunt resistor $R_\text{s}$, and a parasitic resistance $R_\text{p}$ (due to wires and contacts) in series with the SQUID. $I_\text{SQUID}$ is measured with a serial SQUID array amplifier (SSAA) which is operated in FLL mode. A control current $I_\text{ctrl}$ is applied via the voltage $V_\text{ctrl}$ and resistor $R_\text{ctrl}$. The external magnetic field $B_\text{a}$ is always applied perpendicular to the plane of the SQUID loop.

Figure \ref{fig3}(b) shows $I_\text{SQUID}(V_\text{b})$ of the 3JJ-SoL milled by Ne-FIB for different  applied magnetic fields $B_\text{a}$, with $I_\text{ctrl}=0$. For $|I_\text{SQUID}|<I_\text{c}$ the curves are linear, with an inverse slope $V_\text{b}/I_\text{SQUID}\approx R_\text{b}(1+R_\text{p}/R_\text{s})$ (in the limit $R_\text{b}\gg R_\text{s},R_\text{p}$). From the measured $V_\text{b}/I_\text{SQUID}=8.06~\text{k}\Omega$, we extract $R_\text{p}=0.3~\Omega$. At the kinks in $I_\text{SQUID}(V_\text{b})$ the critical current is reached.
Figure \ref{fig3}(c) shows, for both 3JJ-SoLs (at $I_\text{ctrl}=0$), the $I_\text{c}(B)$ curves that have been extracted from $I_\text{SQUID}(V_\text{b})$ data. The central maxima of the $I_\text{c}(B)$ curves are shifted along the field axis, as expected from the asymmetry of a SQUID with three JJs~\cite{meltzer_multi-terminal_2016}. The neighboring maxima are significantly suppressed for the 3JJ-SoL milled by Ne-FIB  or almost absent for the 3JJ-SoL milled by He-FIB SQUID due to the decay of $I_\text{c}$ as $B$ approaches the upper critical field of the Nb film. This effect is a consequence of the small size of the SQUID and the resulting large SQUID oscillation period (equivalent to $\Phi_0$), which corresponds to the difference $\Delta B\approx 0.42~$T between the minima in $I_\text{c}(B)$ for the Ne-FIB SQUID. This period, in turn, yields an effective area $A_\text{eff}=\Phi_0/\Delta B\approx 4.92\times 10^3~\text{nm}^2$ corresponding to an effective diameter $d_\text{eff}=(A_\text{eff}\cdot 4/\pi)^{\nicefrac{1}{2}}\approx 80~$nm. For the 3JJ-SoL milled by He-FIB we find $\Delta B \approx 0.70$~T, resulting in  $d_\text{eff}\approx 60~$nm. For the 3JJ-SoL milled by Ne-FIB, we also show in Fig.~\ref{fig3}(c) the $I_\text{c}(B)$ curve obtained with $I_\text{ctrl}=30~$\textmu A, which is shifted by approximately $-30~$mT along the $B$ axis. This demonstrates an important feature of SQUIDs with three JJs: they provide the possibility to maintain the optimum working point in variable magnetic fields by adjusting $I_\text{ctrl}$.

At selected working points ($V_\text{b}$, $V_\text{ctrl}$) we measured the root-mean-square (rms) spectral density of current noise $S_{I_\text{SQUID}}^{\nicefrac{1}{2}}$ with the SSAA. Combined with the SQUID magnetic responses (transfer functions) $\partial I_\text{SQ}/\partial\Phi$ and $\partial I_\text{SQ} /\partial B$, we determine the respective rms flux noise $S_\Phi^{\nicefrac{1}{2}}=S_{I_\text{SQUID}}^{\nicefrac{1}{2}}/(\partial I_\text{SQ}/\partial\Phi)$ and field noise $S_B^{\nicefrac{1}{2}}=S_{I_\text{SQUID}}^{\nicefrac{1}{2}}/(\partial I_\text{SQ}/\partial B)$. Figure \ref{fig3}(d) shows measurements in the white noise limit at 12~kHz of $S_\Phi^{\nicefrac{1}{2}}$ (also converted into $S_B^{\nicefrac{1}{2}}$) vs applied field for the 3JJ-SoL milled by Ne-FIB. We find a pronounced minimum in the flux noise of $S_\Phi^{\nicefrac{1}{2}} \approx  0.3~$\textmu $\Phi_0/\sqrt{\text{Hz}}$, corresponding to a field noise of $S_B^{\nicefrac{1}{2}} \approx 120 ~\text{nT}/\sqrt{\text{Hz}}$. For $I_\text{ctrl}=0$ the minimum in noise vs $B$ is at $\sim 160~$mT. With $I_\text{ctrl}=30~$\textmu A, we can shift the noise minimum to $\sim 130~$mT. This again demonstrates the benefit of SQUIDs with three JJs: they provide the possibility to maintain optimum sensitivity in variable magnetic fields via the adjustment of $I_\text{ctrl}$, which can be controlled by a PLL. Importantly, the amplitudes of $I_\text{ctrl}$ required to shift the optimum working point in a magnetic field are about an order of magnitude lower that what is required for $I_\text{mod}$ for flux feedback in the case of a SQUID with two JJs. The smaller resulting stray fields make imaging with a 3JJ-SoL in a PLL less invasive than imaging with a 2JJ-SoL in a FLL.

\section{Magnetic imaging}\label{sec4}

In order to demonstrate the spatial resolution of magnetic microscopy with these SQUID-on-lever probes, we map the stray magnetic field at the surface of bulk \CuOSeO. At low temperature, this insulating cubic helimagnet hosts a number of modulated magnetic phases, including low-temperature skyrmion (LTS) and helical (H) phases, which produce nanometer-scale magnetic field patterns at the sample surface. In particular, in magnetic fields applied along $\langle 100 \rangle$, previous scanning SQUID microscopy has shown that the LTS phase appears on the corresponding $\{100\}$ surface in the form of clusters of disordered skyrmions within a field-polarized (FP) background~\cite{marchiori_imaging_2024}. In images of the magnetic field, individual skyrmions generate a point-like reduction in the out-of-plane stray-field, as a result of their core magnetization opposing the surrounding FP phase. Because of their small size -- according to micromagnetic simulations skyrmions in \CuOSeO~have a radius of 10~nm to the region where their out-of-plane magnetization vanishes -- these features are ideal for calibrating the point spread function of our scanning probe and determining its spatial resolution.

\begin{figure*}[t]%
\centering
\includegraphics[width=1\textwidth]{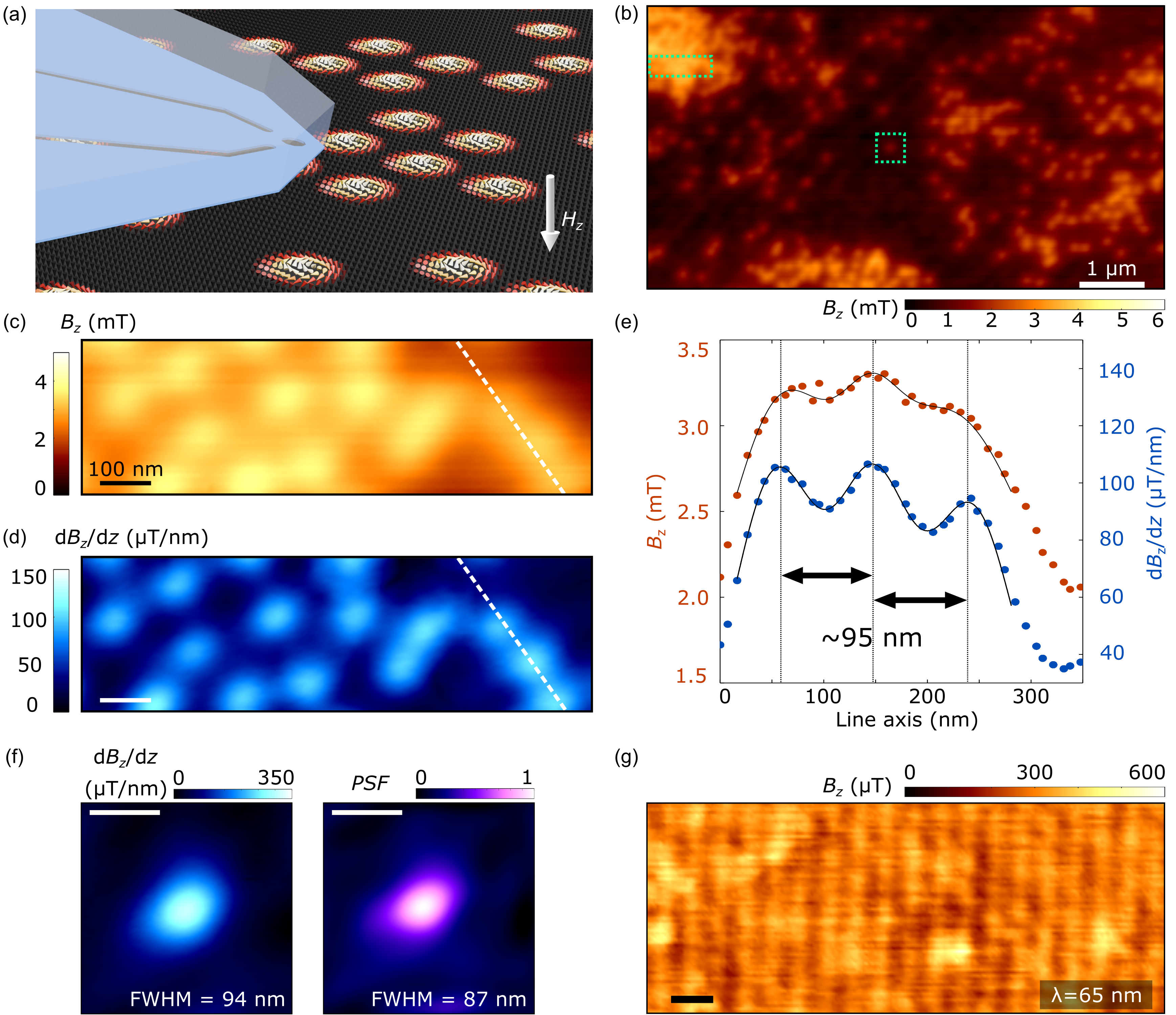}
\caption{Magnetic field microscopy. (a) Schematic of SQUID-on-lever scanning probe over a skyrmion.
(b) $B_z(x,y)$ measured $35 \pm 10$~nm above the surface of \CuOSeO in the low-temperature skyrmion (LTS) phase at $T = 5$~K. Green dotted frames indicate zoomed in regions in (c) and (d).
(c), (d) $B_z (x,y)$ and $\dBdz (x,y)$ of skyrmion clusters. The scale bars of this image and all following ones correspond to 100~nm. 
(e) Line cuts corresponding to the white dotted lines indicated in (c) and (d). Points indicate measured data, while lines represent Gaussian fits.
(f) $\dBdz (x,y)$ and $PSF$ of the SoL extracted from the measurement of a single skyrmion with 2D Gaussian fit FWHM. 
(g) $B_z (x,y)$ of the in-plane helical phase. Fourier space analysis yields a periodic spacing of $65 \pm 5$~nm.}
\label{fig4}
\end{figure*}

Figure \ref{fig4}(a) shows a schematic of the SQUID-on-lever probe above the magnetization texture of a cluster of skyrmions within a FP background. In order to enter the LTS phase, we initially saturate the system with an out-of-plane applied magnetic field along $[001]$ of $\mu_0 H_z = 200$~mT at $T = 5$~K. We then reverse this field, first applying $-150$~mT, and map the stray magnetic field just above the $(001)$ surface of the crystal.  We measure both the component of the stray field $B_z$ perpendicular to the surface and, by oscillating the sample along the $z$-axis, we also measure $dB_z/dz$.  We image the near-surface stray field as a function of increasing reverse field. At $-155$~mT, images show the coexistence of domains of tilted spiral phase, FP phase, and clusters of LTS phase, 
as expected from previous measurements~\cite{marchiori_imaging_2024}.  At $-160$~mT, only clusters of skyrmions remain in an FP background, as shown in Fig.~\ref{fig4}(b). 

Figures \ref{fig4}(c) and (d) focus on a dense cluster of skyrmions, highlighted by the green rectangle in (b), showing $B_z(x,y)$ and $dB_z/dz (x,y)$, respectively. From the partially overlapping features, the superior resolution of the $\dBdz (x,y)$ image compared to the $B_z(x,y)$ image is clear. This improvement results from a combination of both the higher spatial resolution characteristic of measuring a magnetic field derivative compared to a magnetic field~\cite{marchiori_nanoscale_2022} and the noise reduction offered by the spectral filtering of the lock-in-based derivative measurement. Line cuts through the images along the white dotted lines, shown in Fig.~\ref{fig4}(e), cut across a row of three nearby skyrmions. These skyrmions, which are separated by about 95~nm, are resolved in both the field and field-derivative measurements, although their separation in the latter is more evident. A common definition of spatial resolution is the smallest distance between two point-like objects for which each appears distinct. Under the Sparrow criterion, this distance can be defined as the separation below which the minimum between two point-like features disappears.  Given the clearly defined minima in the field derivative measurement and the fact that the stray field produced by the individual skyrmions in the scanning plane is not point-like, these images demonstrate that the SQUID-on-lever is capable of spatial resolution well below 100~nm.

Figure \ref{fig4}(f) shows the contrast of a single skyrmion, highlighted by the green square in (b) and measured at a constant tip-sample spacing of $35 \pm 10$~nm. By fitting the $B_z(x,y)$ and $\dBdz (x,y)$ profiles with Gaussian functions, we find a full width at half maximum (FWHM) of 130 and 94~nm, respectively (only $\dBdz (x,y)$ is shown in Fig.~\ref{fig4}(f)). Given that the SQUID does not measure magnetic field at each point in space, but rather measures the integrated magnetic flux threading through its diameter, its response to a spatially varying magnetic field is characterized by a point spread function (PSF). Slight asymmetries in this PSF, resulting from the shape of the particular SoL sensor, likely result in the asymmetry visible in the images of the skyrmion, whose field profile should be circularly symmetric.

In order to determine the PSF of our SQUID-on-lever probe, we compare $\dBdz (x,y)$ images of a single skyrmion measured by the SQUID, to the idealized stray field expected for a skyrmion in \CuOSeO.  We simulate this stray-field above the surface of the sample using the micromagnetic software package \textit{MuMax3}, which is based on the Landau-Liftshitz-Gilbert formalism~\cite{vansteenkiste_design_2014,exl_labontes_2014}.  We then use an iterative Landweber algorithm to extract a PSF from the simulated and measured stray field images. The algorithm starts from an initial estimate of the PSF, which after each step is successively improved upon  by reducing the error between the measured stray field image and that generated by convolving the simulated stray-field with the PSF. The algorithm converges once this error is comparable to the measurement noise \cite{combettes_proximal_2011}.
The PSF extracted from the $\dBdz (x,y)$ map of the single skryrmion in the first panel of Fig.~\ref{fig4}(f) is shown in the second panel.  This point spread function has a FWHM of 87~nm, giving another measure of the probe's spatial resolution. Given the similar size of the point spread function and the measured $\dBdz (x,y)$ profile of the single skyrmion, it is clear that the field profile of the skyrmion itself at this tip-sample spacing is much smaller than the resolution of our probe.  In that sense, the images in Fig.~\ref{fig4}(f) reveal the size and asymmetries of the SQUID-on-lever probe rather than those of the skyrmion's stray field.

As a further test of the probe's response to small magnetic features, we also image the short-wavelength magnetic modulation produced by an in-plane H phase.  At low temperature and in low applied magnetic fields, \CuOSeO\ can enter a multi-domain H phase, in which domains of magnetization helices, which propagate along $[100]$, $[010]$, or $[001]$, coexist.  At a $\{100\}$ surface, in-plane H domains appear as sinusoidal stray field modulations with a period given by the wavelength of the intrinsic helimagnetic order in \CuOSeO, $\lambda_m = 62$~nm~\cite{adams_long-wavelength_2012,marchiori_imaging_2024}. In order to produce in-plane H domains at the sample surface, we apply an in-plane magnetic field  $\mu_0 H_x = 200$~mT aligned approximately along $[100]$ to fully polarize the sample at $T = 5$~K.  We then decrease the applied field down to less than 12~mT and make maps of $B_z (x,y)$ above the surface. A characteristic image is shown in Fig.~\ref{fig4}(g), revealing a sinusoidal stray field modulation with a period of $65 \pm 5$~nm propagating along $[100]$. The error in the measurement of the period is due to a systematic uncertainty in our calibration of the piezo-electric scanner used to scan the sample.  Apart from non-periodic modulations in the $B_z (x,y)$ maps, due to surface roughness, the images are consistent with behavior expected from the stray field at the surface of an in-plane H phase.  
The clarity of the contrast provides a final confirmation of the SQUID-on-lever's sensitivity to spatial variations in magnetic field on length-scales smaller than 100 nm.

\section{Conclusion}\label{sec5}

We demonstrate a method for fabricating robust SQUID-on-lever probes for high-sensitivity SSM. The probes combine the advantages of previously developed SSM probes, including the possibility of operating in magnetic fields exceeding 0.5~T, spatial resolution below 100~nm, the possibility of integrating on-tip circuitry -- such as a modulation line or additional JJs -- and the ability to fabricate large quantities of probes. The spatial resolution is calibrated against nanometer-scale magnetization textures, yielding a best-case point-spread function with a FWHM of 87~nm.  Images of a modulated magnetization pattern with a period of 65~nm further demonstrate the probe's sensitivity to magnetic field variations with short spatial wavelengths.  

SQUID-on-lever probes with integrated modulation lines or control lines could be operated in a flux-locked or phase-locked loop mode to extend their range of sensitivity in magnetic field, reducing the ``blind spots'' intrinsic to SQUID-on-tip sensors~\cite{uri_electrically_2016}.  All of these properties, combined with the ease-of-use of the cantilever platform and the robustness of the probes will dramatically expand the applications of nanometer-scale SSM in the fields of magnetism, superconductivity, and low-temperature transport.


\begin{acknowledgments}
We acknowledge support by the European Commission under H2020 FET Open grant “FIBsuperProbes” (Grant No. 892427) and the Swiss National Science Foundation under Grant No. 200020-207933. We also gratefully acknowledge support by the COST actions NANOCOHYBRI (CA16218), FIT4NANO (CA19140) and SUPERQUMAP (CA21144).
\end{acknowledgments}

\section*{Data Availability}

The data that support the findings of this study are available from the corresponding author upon reasonable request.

\appendix

\section{\texorpdfstring{$\text{Cu}_2 \text{OSeO}_3$}{Cu₂OSeO₃} sample}

The $\text{Cu}_2 \text{OSeO}_3$ crystal is grown via chemical vapor transport with CuO and SeO$_2$ as starting materials and HCl as the transport agent~\cite{baral_tuning_2022}. A well-faceted single crystal is chosen with a natural $\{100\}$ facet. The orientation is verified using a Laue camera.  The facet is then mechanically polished and the orientation is subsequently re-verified using the Laue camera.  The miscut angle is measured using rocking curves with x-ray diffraction to be less than $1^\circ$.

\section{Scanning SQUID microscopy}

We perform magnetic imaging in a custom-built microscope under high vacuum inside a \(^4\text{He}\) cryostat at 4.2~K. 
The microscope contains a vector-magnet that can apply a field of up to 1~T either in- or out-of-plane, i.e.\ along $\hat{x}$ or $\hat{z}$.
The SQUID-on-lever scanning probe can perform both non-contact AFM and SSM.
At 4.2~K and in high vacuum, the cantilever has a resonance frequency of 577~kHz and a mechanical quality factor of $~15000$. Upon approaching the sample, the cantilever's resonance frequency shifts at a distance of more than 500~nm, allowing for excellent tip-sample distance control. The cantilever is excited to an amplitude of $10$~nm at its fundamental mechanical resonance frequency by a piezo-electric actuator driven by a phase-locked loop. 
Displacement oscillations are detected using a fiber-optic interferometer~\cite{wyss_magnetic_2022}.
The SQUID is characterized and operated in a semi-voltage-biased circuit, in which the current response $I_{\text{SQUID}}$ is measured by a series SQUID array amplifier (Magnicon). 
We calibrate the SQUID response to magnetic field by measuring ($I_\text{SQUID}$ vs $\mu_0H_z$) before and after each scan in a field range significantly larger than the field created by the \CuOSeO.
This response provides a measure of the local magnetic field threading through the SQUID, i.e. the out-of plane component of the magnetic field $B_z (x,y)$.
We further oscillate the sample out-of-plane at a frequency of 177~Hz with amplitude $\Delta z\approx 1.5$~nm and use a lock-in amplifier to detect the AC field $B_\text{ac}$ in a narrow bandwidth.
To first-order, the AC field is given by $B_\text{ac} \approx \Delta z \cdot \dBdz$ and thus provides a measure of the out-of-plane field gradient threading the SQUID.

\section{Micromagnetic simulations}
We use the finite element software package \textit{MuMax3} \cite{vansteenkiste_design_2014, exl_labontes_2014}, which is based on the Landau-Lifshitz-Gilbert formalism, to simulate the magnetization profile of individual skyrmions in $\text{Cu}_2 \text{OSeO}_3$. As simulation parameters we use a saturation magnetization of \SI{103}{\kilo\ampere/\meter}, an exchange stiffness of \SI{0.35}{\pico\joule/\meter}, a bulk Dzyaloshinskii-Moriya constant of \SI{0.74}{\micro\joule/\meter\squared}, and a cubic anisotropy constant of \SI{1.2}{\kilo\joule/\meter\cubed}. Cubic anisotropy axes point along the axes of the $xyz$ coordinate system.
To mimic the geometry of a bulk piece of single-crystal $\text{Cu}_2 \text{OSeO}_3$, we employ periodic boundary conditions in $x$ and $y$ with 5 repetitions on each side of a simulation volume of \SI{1008}{\nano\meter} $\times$ \SI{1008}{\nano\meter} $\times$ \SI{200}{\nano\meter}. The finite element mesh is set to a size of 240 $\times$ 240 $\times$ 40 cells with corresponding cell sizes of \SI{4.2}{\nano\meter} $\times$ \SI{4.2}{\nano\meter} $\times$ \SI{5}{\nano\meter}, respectively.
To facilitate stable appearance of individual skyrmions, we apply a constant, homogeneous magnetic field in $z$ with a strength of \SI{0.2}{\tesla}, and initiate the magnetization with a Bloch skyrmion, which has its core pointing along $z$ in the center of the simulation volume. We then relax the system to the next energetic minimum that is found by the solver.

\nocite{*}

\bibliography{HeNeSoL}

\begin{thebibliography}{48}%
\makeatletter
\providecommand \@ifxundefined [1]{%
 \@ifx{#1\undefined}
}%
\providecommand \@ifnum [1]{%
 \ifnum #1\expandafter \@firstoftwo
 \else \expandafter \@secondoftwo
 \fi
}%
\providecommand \@ifx [1]{%
 \ifx #1\expandafter \@firstoftwo
 \else \expandafter \@secondoftwo
 \fi
}%
\providecommand \natexlab [1]{#1}%
\providecommand \enquote  [1]{``#1''}%
\providecommand \bibnamefont  [1]{#1}%
\providecommand \bibfnamefont [1]{#1}%
\providecommand \citenamefont [1]{#1}%
\providecommand \href@noop [0]{\@secondoftwo}%
\providecommand \href [0]{\begingroup \@sanitize@url \@href}%
\providecommand \@href[1]{\@@startlink{#1}\@@href}%
\providecommand \@@href[1]{\endgroup#1\@@endlink}%
\providecommand \@sanitize@url [0]{\catcode `\\12\catcode `\$12\catcode
  `\&12\catcode `\#12\catcode `\^12\catcode `\_12\catcode `\%12\relax}%
\providecommand \@@startlink[1]{}%
\providecommand \@@endlink[0]{}%
\providecommand \url  [0]{\begingroup\@sanitize@url \@url }%
\providecommand \@url [1]{\endgroup\@href {#1}{\urlprefix }}%
\providecommand \urlprefix  [0]{URL }%
\providecommand \Eprint [0]{\href }%
\providecommand \doibase [0]{https://doi.org/}%
\providecommand \selectlanguage [0]{\@gobble}%
\providecommand \bibinfo  [0]{\@secondoftwo}%
\providecommand \bibfield  [0]{\@secondoftwo}%
\providecommand \translation [1]{[#1]}%
\providecommand \BibitemOpen [0]{}%
\providecommand \bibitemStop [0]{}%
\providecommand \bibitemNoStop [0]{.\EOS\space}%
\providecommand \EOS [0]{\spacefactor3000\relax}%
\providecommand \BibitemShut  [1]{\csname bibitem#1\endcsname}%
\let\auto@bib@innerbib\@empty
\bibitem [{\citenamefont {Christensen}\ \emph {et~al.}(2024)\citenamefont
  {Christensen}, \citenamefont {Staub}, \citenamefont {Devidas}, \citenamefont
  {Kalisky}, \citenamefont {Nowack}, \citenamefont {Webb}, \citenamefont
  {Andersen}, \citenamefont {Huck}, \citenamefont {Broadway}, \citenamefont
  {Wagner}, \citenamefont {Maletinsky}, \citenamefont {Sar}, \citenamefont
  {Du}, \citenamefont {Yacoby}, \citenamefont {Collomb}, \citenamefont
  {Bending}, \citenamefont {Oral}, \citenamefont {Hug}, \citenamefont {Mandru},
  \citenamefont {Neu}, \citenamefont {Schumacher}, \citenamefont {Sievers},
  \citenamefont {Saito}, \citenamefont {Khajetoorians}, \citenamefont
  {Hauptmann}, \citenamefont {Baumann}, \citenamefont {Eichler}, \citenamefont
  {Degen}, \citenamefont {McCord}, \citenamefont {Vogel}, \citenamefont
  {Fiebig}, \citenamefont {Fischer}, \citenamefont {Hierro-Rodriguez},
  \citenamefont {Finizio}, \citenamefont {Dhesi}, \citenamefont {Donnelly},
  \citenamefont {Büttner}, \citenamefont {Kfir}, \citenamefont {Hu},
  \citenamefont {Zayko}, \citenamefont {Eisebitt}, \citenamefont {Pfau},
  \citenamefont {Frömter}, \citenamefont {Kläui}, \citenamefont {Yasin},
  \citenamefont {McMorran}, \citenamefont {Seki}, \citenamefont {Yu},
  \citenamefont {Lubk}, \citenamefont {Wolf}, \citenamefont {Pryds},
  \citenamefont {Makarov},\ and\ \citenamefont
  {Poggio}}]{christensen_2024_2024}%
  \BibitemOpen
  \bibfield  {author} {\bibinfo {author} {\bibfnamefont {D.~V.}\ \bibnamefont
  {Christensen}}, \bibinfo {author} {\bibfnamefont {U.}~\bibnamefont {Staub}},
  \bibinfo {author} {\bibfnamefont {T.~R.}\ \bibnamefont {Devidas}}, \bibinfo
  {author} {\bibfnamefont {B.}~\bibnamefont {Kalisky}}, \bibinfo {author}
  {\bibfnamefont {K.~C.}\ \bibnamefont {Nowack}}, \bibinfo {author}
  {\bibfnamefont {J.~L.}\ \bibnamefont {Webb}}, \bibinfo {author}
  {\bibfnamefont {U.~L.}\ \bibnamefont {Andersen}}, \bibinfo {author}
  {\bibfnamefont {A.}~\bibnamefont {Huck}}, \bibinfo {author} {\bibfnamefont
  {D.~A.}\ \bibnamefont {Broadway}}, \bibinfo {author} {\bibfnamefont
  {K.}~\bibnamefont {Wagner}}, \bibinfo {author} {\bibfnamefont
  {P.}~\bibnamefont {Maletinsky}}, \bibinfo {author} {\bibfnamefont {T.~v.~d.}\
  \bibnamefont {Sar}}, \bibinfo {author} {\bibfnamefont {C.~R.}\ \bibnamefont
  {Du}}, \bibinfo {author} {\bibfnamefont {A.}~\bibnamefont {Yacoby}}, \bibinfo
  {author} {\bibfnamefont {D.}~\bibnamefont {Collomb}}, \bibinfo {author}
  {\bibfnamefont {S.}~\bibnamefont {Bending}}, \bibinfo {author} {\bibfnamefont
  {A.}~\bibnamefont {Oral}}, \bibinfo {author} {\bibfnamefont {H.~J.}\
  \bibnamefont {Hug}}, \bibinfo {author} {\bibfnamefont {A.-O.}\ \bibnamefont
  {Mandru}}, \bibinfo {author} {\bibfnamefont {V.}~\bibnamefont {Neu}},
  \bibinfo {author} {\bibfnamefont {H.~W.}\ \bibnamefont {Schumacher}},
  \bibinfo {author} {\bibfnamefont {S.}~\bibnamefont {Sievers}}, \bibinfo
  {author} {\bibfnamefont {H.}~\bibnamefont {Saito}}, \bibinfo {author}
  {\bibfnamefont {A.~A.}\ \bibnamefont {Khajetoorians}}, \bibinfo {author}
  {\bibfnamefont {N.}~\bibnamefont {Hauptmann}}, \bibinfo {author}
  {\bibfnamefont {S.}~\bibnamefont {Baumann}}, \bibinfo {author} {\bibfnamefont
  {A.}~\bibnamefont {Eichler}}, \bibinfo {author} {\bibfnamefont {C.~L.}\
  \bibnamefont {Degen}}, \bibinfo {author} {\bibfnamefont {J.}~\bibnamefont
  {McCord}}, \bibinfo {author} {\bibfnamefont {M.}~\bibnamefont {Vogel}},
  \bibinfo {author} {\bibfnamefont {M.}~\bibnamefont {Fiebig}}, \bibinfo
  {author} {\bibfnamefont {P.}~\bibnamefont {Fischer}}, \bibinfo {author}
  {\bibfnamefont {A.}~\bibnamefont {Hierro-Rodriguez}}, \bibinfo {author}
  {\bibfnamefont {S.}~\bibnamefont {Finizio}}, \bibinfo {author} {\bibfnamefont
  {S.~S.}\ \bibnamefont {Dhesi}}, \bibinfo {author} {\bibfnamefont
  {C.}~\bibnamefont {Donnelly}}, \bibinfo {author} {\bibfnamefont
  {F.}~\bibnamefont {Büttner}}, \bibinfo {author} {\bibfnamefont
  {O.}~\bibnamefont {Kfir}}, \bibinfo {author} {\bibfnamefont {W.}~\bibnamefont
  {Hu}}, \bibinfo {author} {\bibfnamefont {S.}~\bibnamefont {Zayko}}, \bibinfo
  {author} {\bibfnamefont {S.}~\bibnamefont {Eisebitt}}, \bibinfo {author}
  {\bibfnamefont {B.}~\bibnamefont {Pfau}}, \bibinfo {author} {\bibfnamefont
  {R.}~\bibnamefont {Frömter}}, \bibinfo {author} {\bibfnamefont
  {M.}~\bibnamefont {Kläui}}, \bibinfo {author} {\bibfnamefont {F.~S.}\
  \bibnamefont {Yasin}}, \bibinfo {author} {\bibfnamefont {B.~J.}\ \bibnamefont
  {McMorran}}, \bibinfo {author} {\bibfnamefont {S.}~\bibnamefont {Seki}},
  \bibinfo {author} {\bibfnamefont {X.}~\bibnamefont {Yu}}, \bibinfo {author}
  {\bibfnamefont {A.}~\bibnamefont {Lubk}}, \bibinfo {author} {\bibfnamefont
  {D.}~\bibnamefont {Wolf}}, \bibinfo {author} {\bibfnamefont {N.}~\bibnamefont
  {Pryds}}, \bibinfo {author} {\bibfnamefont {D.}~\bibnamefont {Makarov}},\
  and\ \bibinfo {author} {\bibfnamefont {M.}~\bibnamefont {Poggio}},\
  }\bibfield  {title} {\bibinfo {title} {2024 roadmap on magnetic microscopy
  techniques and their applications in materials science},\ }\href
  {https://doi.org/10.1088/2515-7639/ad31b5} {\bibfield  {journal} {\bibinfo
  {journal} {Journal of Physics: Materials}\ }\textbf {\bibinfo {volume} {7}},\
  \bibinfo {pages} {032501} (\bibinfo {year} {2024})}\BibitemShut {NoStop}%
\bibitem [{\citenamefont {Vasyukov}\ \emph {et~al.}(2018)\citenamefont
  {Vasyukov}, \citenamefont {Ceccarelli}, \citenamefont {Wyss}, \citenamefont
  {Gross}, \citenamefont {Schwarb}, \citenamefont {Mehlin}, \citenamefont
  {Rossi}, \citenamefont {Tütüncüoglu}, \citenamefont {Heimbach},
  \citenamefont {Zamani}, \citenamefont {Kovács}, \citenamefont {Fontcuberta~i
  Morral}, \citenamefont {Grundler},\ and\ \citenamefont
  {Poggio}}]{vasyukov_imaging_2018}%
  \BibitemOpen
  \bibfield  {author} {\bibinfo {author} {\bibfnamefont {D.}~\bibnamefont
  {Vasyukov}}, \bibinfo {author} {\bibfnamefont {L.}~\bibnamefont
  {Ceccarelli}}, \bibinfo {author} {\bibfnamefont {M.}~\bibnamefont {Wyss}},
  \bibinfo {author} {\bibfnamefont {B.}~\bibnamefont {Gross}}, \bibinfo
  {author} {\bibfnamefont {A.}~\bibnamefont {Schwarb}}, \bibinfo {author}
  {\bibfnamefont {A.}~\bibnamefont {Mehlin}}, \bibinfo {author} {\bibfnamefont
  {N.}~\bibnamefont {Rossi}}, \bibinfo {author} {\bibfnamefont
  {G.}~\bibnamefont {Tütüncüoglu}}, \bibinfo {author} {\bibfnamefont
  {F.}~\bibnamefont {Heimbach}}, \bibinfo {author} {\bibfnamefont {R.~R.}\
  \bibnamefont {Zamani}}, \bibinfo {author} {\bibfnamefont {A.}~\bibnamefont
  {Kovács}}, \bibinfo {author} {\bibfnamefont {A.}~\bibnamefont {Fontcuberta~i
  Morral}}, \bibinfo {author} {\bibfnamefont {D.}~\bibnamefont {Grundler}},\
  and\ \bibinfo {author} {\bibfnamefont {M.}~\bibnamefont {Poggio}},\
  }\bibfield  {title} {\bibinfo {title} {Imaging stray magnetic field of
  individual ferromagnetic nanotubes},\ }\href
  {https://doi.org/10.1021/acs.nanolett.7b04386} {\bibfield  {journal}
  {\bibinfo  {journal} {Nano Letters}\ }\textbf {\bibinfo {volume} {18}},\
  \bibinfo {pages} {964} (\bibinfo {year} {2018})}\BibitemShut {NoStop}%
\bibitem [{\citenamefont {Wyss}\ \emph {et~al.}(2019)\citenamefont {Wyss},
  \citenamefont {Gliga}, \citenamefont {Vasyukov}, \citenamefont {Ceccarelli},
  \citenamefont {Romagnoli}, \citenamefont {Cui}, \citenamefont {Kleibert},
  \citenamefont {Stamps},\ and\ \citenamefont
  {Poggio}}]{wyss_stray-field_2019}%
  \BibitemOpen
  \bibfield  {author} {\bibinfo {author} {\bibfnamefont {M.}~\bibnamefont
  {Wyss}}, \bibinfo {author} {\bibfnamefont {S.}~\bibnamefont {Gliga}},
  \bibinfo {author} {\bibfnamefont {D.}~\bibnamefont {Vasyukov}}, \bibinfo
  {author} {\bibfnamefont {L.}~\bibnamefont {Ceccarelli}}, \bibinfo {author}
  {\bibfnamefont {G.}~\bibnamefont {Romagnoli}}, \bibinfo {author}
  {\bibfnamefont {J.}~\bibnamefont {Cui}}, \bibinfo {author} {\bibfnamefont
  {A.}~\bibnamefont {Kleibert}}, \bibinfo {author} {\bibfnamefont {R.~L.}\
  \bibnamefont {Stamps}},\ and\ \bibinfo {author} {\bibfnamefont
  {M.}~\bibnamefont {Poggio}},\ }\bibfield  {title} {\bibinfo {title}
  {Stray-field imaging of a chiral artificial spin ice during magnetization
  reversal},\ }\href {https://doi.org/10.1021/acsnano.9b05428} {\bibfield
  {journal} {\bibinfo  {journal} {ACS Nano}\ }\textbf {\bibinfo {volume}
  {13}},\ \bibinfo {pages} {13910} (\bibinfo {year} {2019})}\BibitemShut
  {NoStop}%
\bibitem [{\citenamefont {Bert}\ \emph {et~al.}(2011)\citenamefont {Bert},
  \citenamefont {Kalisky}, \citenamefont {Bell}, \citenamefont {Kim},
  \citenamefont {Hikita}, \citenamefont {Hwang},\ and\ \citenamefont
  {Moler}}]{bert_direct_2011}%
  \BibitemOpen
  \bibfield  {author} {\bibinfo {author} {\bibfnamefont {J.~A.}\ \bibnamefont
  {Bert}}, \bibinfo {author} {\bibfnamefont {B.}~\bibnamefont {Kalisky}},
  \bibinfo {author} {\bibfnamefont {C.}~\bibnamefont {Bell}}, \bibinfo {author}
  {\bibfnamefont {M.}~\bibnamefont {Kim}}, \bibinfo {author} {\bibfnamefont
  {Y.}~\bibnamefont {Hikita}}, \bibinfo {author} {\bibfnamefont {H.~Y.}\
  \bibnamefont {Hwang}},\ and\ \bibinfo {author} {\bibfnamefont {K.~A.}\
  \bibnamefont {Moler}},\ }\bibfield  {title} {\bibinfo {title} {Direct imaging
  of the coexistence of ferromagnetism and superconductivity at the
  {LaAlO}$_{\textrm{3}}$ /{SrTiO}$_{\textrm{3}}$ interface},\ }\href
  {https://doi.org/10.1038/nphys2079} {\bibfield  {journal} {\bibinfo
  {journal} {Nature Physics}\ }\textbf {\bibinfo {volume} {7}},\ \bibinfo
  {pages} {767} (\bibinfo {year} {2011})}\BibitemShut {NoStop}%
\bibitem [{\citenamefont {Christensen}\ \emph {et~al.}(2019)\citenamefont
  {Christensen}, \citenamefont {Frenkel}, \citenamefont {Chen}, \citenamefont
  {Xie}, \citenamefont {Chen}, \citenamefont {Hikita}, \citenamefont {Smith},
  \citenamefont {Klein}, \citenamefont {Hwang}, \citenamefont {Pryds},\ and\
  \citenamefont {Kalisky}}]{christensen_strain-tunable_2019}%
  \BibitemOpen
  \bibfield  {author} {\bibinfo {author} {\bibfnamefont {D.~V.}\ \bibnamefont
  {Christensen}}, \bibinfo {author} {\bibfnamefont {Y.}~\bibnamefont
  {Frenkel}}, \bibinfo {author} {\bibfnamefont {Y.~Z.}\ \bibnamefont {Chen}},
  \bibinfo {author} {\bibfnamefont {Y.~W.}\ \bibnamefont {Xie}}, \bibinfo
  {author} {\bibfnamefont {Z.~Y.}\ \bibnamefont {Chen}}, \bibinfo {author}
  {\bibfnamefont {Y.}~\bibnamefont {Hikita}}, \bibinfo {author} {\bibfnamefont
  {A.}~\bibnamefont {Smith}}, \bibinfo {author} {\bibfnamefont
  {L.}~\bibnamefont {Klein}}, \bibinfo {author} {\bibfnamefont {H.~Y.}\
  \bibnamefont {Hwang}}, \bibinfo {author} {\bibfnamefont {N.}~\bibnamefont
  {Pryds}},\ and\ \bibinfo {author} {\bibfnamefont {B.}~\bibnamefont
  {Kalisky}},\ }\bibfield  {title} {\bibinfo {title} {Strain-tunable magnetism
  at oxide domain walls},\ }\href {https://doi.org/10.1038/s41567-018-0363-x}
  {\bibfield  {journal} {\bibinfo  {journal} {Nature Physics}\ }\textbf
  {\bibinfo {volume} {15}},\ \bibinfo {pages} {269} (\bibinfo {year}
  {2019})}\BibitemShut {NoStop}%
\bibitem [{\citenamefont {Wang}\ \emph {et~al.}(2015)\citenamefont {Wang},
  \citenamefont {Li}, \citenamefont {Lü}, \citenamefont {Paudel},
  \citenamefont {Leusink}, \citenamefont {Hoek}, \citenamefont {Poccia},
  \citenamefont {Vailionis}, \citenamefont {Venkatesan}, \citenamefont {Coey},
  \citenamefont {Tsymbal}, \citenamefont {Ariando},\ and\ \citenamefont
  {Hilgenkamp}}]{wang_imaging_2015}%
  \BibitemOpen
  \bibfield  {author} {\bibinfo {author} {\bibfnamefont {X.~R.}\ \bibnamefont
  {Wang}}, \bibinfo {author} {\bibfnamefont {C.~J.}\ \bibnamefont {Li}},
  \bibinfo {author} {\bibfnamefont {W.~M.}\ \bibnamefont {Lü}}, \bibinfo
  {author} {\bibfnamefont {T.~R.}\ \bibnamefont {Paudel}}, \bibinfo {author}
  {\bibfnamefont {D.~P.}\ \bibnamefont {Leusink}}, \bibinfo {author}
  {\bibfnamefont {M.}~\bibnamefont {Hoek}}, \bibinfo {author} {\bibfnamefont
  {N.}~\bibnamefont {Poccia}}, \bibinfo {author} {\bibfnamefont
  {A.}~\bibnamefont {Vailionis}}, \bibinfo {author} {\bibfnamefont
  {T.}~\bibnamefont {Venkatesan}}, \bibinfo {author} {\bibfnamefont {J.~M.~D.}\
  \bibnamefont {Coey}}, \bibinfo {author} {\bibfnamefont {E.~Y.}\ \bibnamefont
  {Tsymbal}}, \bibinfo {author} {\bibnamefont {Ariando}},\ and\ \bibinfo
  {author} {\bibfnamefont {H.}~\bibnamefont {Hilgenkamp}},\ }\bibfield  {title}
  {\bibinfo {title} {Imaging and control of ferromagnetism in
  {LaMnO}$_{\textrm{3}}$/{SrTiO}$_{\textrm{3}}$ heterostructures},\ }\href
  {https://doi.org/10.1126/science.aaa5198} {\bibfield  {journal} {\bibinfo
  {journal} {Science}\ }\textbf {\bibinfo {volume} {349}},\ \bibinfo {pages}
  {716} (\bibinfo {year} {2015})}\BibitemShut {NoStop}%
\bibitem [{\citenamefont {Lachman}\ \emph {et~al.}(2015)\citenamefont
  {Lachman}, \citenamefont {Young}, \citenamefont {Richardella}, \citenamefont
  {Cuppens}, \citenamefont {Naren}, \citenamefont {Anahory}, \citenamefont
  {Meltzer}, \citenamefont {Kandala}, \citenamefont {Kempinger}, \citenamefont
  {Myasoedov}, \citenamefont {Huber}, \citenamefont {Samarth},\ and\
  \citenamefont {Zeldov}}]{lachman_visualization_2015}%
  \BibitemOpen
  \bibfield  {author} {\bibinfo {author} {\bibfnamefont {E.~O.}\ \bibnamefont
  {Lachman}}, \bibinfo {author} {\bibfnamefont {A.~F.}\ \bibnamefont {Young}},
  \bibinfo {author} {\bibfnamefont {A.}~\bibnamefont {Richardella}}, \bibinfo
  {author} {\bibfnamefont {J.}~\bibnamefont {Cuppens}}, \bibinfo {author}
  {\bibfnamefont {H.~R.}\ \bibnamefont {Naren}}, \bibinfo {author}
  {\bibfnamefont {Y.}~\bibnamefont {Anahory}}, \bibinfo {author} {\bibfnamefont
  {A.~Y.}\ \bibnamefont {Meltzer}}, \bibinfo {author} {\bibfnamefont
  {A.}~\bibnamefont {Kandala}}, \bibinfo {author} {\bibfnamefont
  {S.}~\bibnamefont {Kempinger}}, \bibinfo {author} {\bibfnamefont
  {Y.}~\bibnamefont {Myasoedov}}, \bibinfo {author} {\bibfnamefont {M.~E.}\
  \bibnamefont {Huber}}, \bibinfo {author} {\bibfnamefont {N.}~\bibnamefont
  {Samarth}},\ and\ \bibinfo {author} {\bibfnamefont {E.}~\bibnamefont
  {Zeldov}},\ }\bibfield  {title} {\bibinfo {title} {Visualization of
  superparamagnetic dynamics in magnetic topological insulators},\ }\href
  {https://doi.org/10.1126/sciadv.1500740} {\bibfield  {journal} {\bibinfo
  {journal} {Science Advances}\ }\textbf {\bibinfo {volume} {1}},\ \bibinfo
  {pages} {e1500740} (\bibinfo {year} {2015})}\BibitemShut {NoStop}%
\bibitem [{\citenamefont {Noah}\ \emph {et~al.}(2023)\citenamefont {Noah},
  \citenamefont {Zur}, \citenamefont {Fridman}, \citenamefont {Singh},
  \citenamefont {Gutfreund}, \citenamefont {Herrera}, \citenamefont {Vakahi},
  \citenamefont {Remennik}, \citenamefont {Huber}, \citenamefont {Gazit},
  \citenamefont {Suderow}, \citenamefont {Steinberg}, \citenamefont {Millo},\
  and\ \citenamefont {Anahory}}]{noah_nano-patterned_2023}%
  \BibitemOpen
  \bibfield  {author} {\bibinfo {author} {\bibfnamefont {A.}~\bibnamefont
  {Noah}}, \bibinfo {author} {\bibfnamefont {Y.}~\bibnamefont {Zur}}, \bibinfo
  {author} {\bibfnamefont {N.}~\bibnamefont {Fridman}}, \bibinfo {author}
  {\bibfnamefont {S.}~\bibnamefont {Singh}}, \bibinfo {author} {\bibfnamefont
  {A.}~\bibnamefont {Gutfreund}}, \bibinfo {author} {\bibfnamefont
  {E.}~\bibnamefont {Herrera}}, \bibinfo {author} {\bibfnamefont
  {A.}~\bibnamefont {Vakahi}}, \bibinfo {author} {\bibfnamefont
  {S.}~\bibnamefont {Remennik}}, \bibinfo {author} {\bibfnamefont {M.~E.}\
  \bibnamefont {Huber}}, \bibinfo {author} {\bibfnamefont {S.}~\bibnamefont
  {Gazit}}, \bibinfo {author} {\bibfnamefont {H.}~\bibnamefont {Suderow}},
  \bibinfo {author} {\bibfnamefont {H.}~\bibnamefont {Steinberg}}, \bibinfo
  {author} {\bibfnamefont {O.}~\bibnamefont {Millo}},\ and\ \bibinfo {author}
  {\bibfnamefont {Y.}~\bibnamefont {Anahory}},\ }\bibfield  {title} {\bibinfo
  {title} {Nano-patterned magnetic edges in {CrGeTe}$_{\textrm{3}}$ for quasi
  1-{D} spintronic devices},\ }\href {https://doi.org/10.1021/acsanm.3c01008}
  {\bibfield  {journal} {\bibinfo  {journal} {ACS Applied Nano Materials}\
  }\textbf {\bibinfo {volume} {6}},\ \bibinfo {pages} {8627} (\bibinfo {year}
  {2023})}\BibitemShut {NoStop}%
\bibitem [{\citenamefont {Zur}\ \emph {et~al.}(2023)\citenamefont {Zur},
  \citenamefont {Noah}, \citenamefont {Boix‐Constant}, \citenamefont
  {Mañas‐Valero}, \citenamefont {Fridman}, \citenamefont {Rama‐Eiroa},
  \citenamefont {Huber}, \citenamefont {Santos}, \citenamefont {Coronado},\
  and\ \citenamefont {Anahory}}]{zur_magnetic_2023}%
  \BibitemOpen
  \bibfield  {author} {\bibinfo {author} {\bibfnamefont {Y.}~\bibnamefont
  {Zur}}, \bibinfo {author} {\bibfnamefont {A.}~\bibnamefont {Noah}}, \bibinfo
  {author} {\bibfnamefont {C.}~\bibnamefont {Boix‐Constant}}, \bibinfo
  {author} {\bibfnamefont {S.}~\bibnamefont {Mañas‐Valero}}, \bibinfo
  {author} {\bibfnamefont {N.}~\bibnamefont {Fridman}}, \bibinfo {author}
  {\bibfnamefont {R.}~\bibnamefont {Rama‐Eiroa}}, \bibinfo {author}
  {\bibfnamefont {M.~E.}\ \bibnamefont {Huber}}, \bibinfo {author}
  {\bibfnamefont {E.~J.~G.}\ \bibnamefont {Santos}}, \bibinfo {author}
  {\bibfnamefont {E.}~\bibnamefont {Coronado}},\ and\ \bibinfo {author}
  {\bibfnamefont {Y.}~\bibnamefont {Anahory}},\ }\bibfield  {title} {\bibinfo
  {title} {Magnetic imaging and domain nucleation in {CrSBr} down to the {2D}
  limit},\ }\href {https://doi.org/10.1002/adma.202307195} {\bibfield
  {journal} {\bibinfo  {journal} {Advanced Materials}\ }\textbf {\bibinfo
  {volume} {35}},\ \bibinfo {pages} {2307195} (\bibinfo {year}
  {2023})}\BibitemShut {NoStop}%
\bibitem [{\citenamefont {Vervelaki}\ \emph {et~al.}(2024)\citenamefont
  {Vervelaki}, \citenamefont {Bagani}, \citenamefont {Jetter}, \citenamefont
  {Doan}, \citenamefont {Chau}, \citenamefont {Gross}, \citenamefont
  {Christensen}, \citenamefont {Bøggild},\ and\ \citenamefont
  {Poggio}}]{vervelaki_visualizing_2024}%
  \BibitemOpen
  \bibfield  {author} {\bibinfo {author} {\bibfnamefont {A.}~\bibnamefont
  {Vervelaki}}, \bibinfo {author} {\bibfnamefont {K.}~\bibnamefont {Bagani}},
  \bibinfo {author} {\bibfnamefont {D.}~\bibnamefont {Jetter}}, \bibinfo
  {author} {\bibfnamefont {M.-H.}\ \bibnamefont {Doan}}, \bibinfo {author}
  {\bibfnamefont {T.~K.}\ \bibnamefont {Chau}}, \bibinfo {author}
  {\bibfnamefont {B.}~\bibnamefont {Gross}}, \bibinfo {author} {\bibfnamefont
  {D.~V.}\ \bibnamefont {Christensen}}, \bibinfo {author} {\bibfnamefont
  {P.}~\bibnamefont {Bøggild}},\ and\ \bibinfo {author} {\bibfnamefont
  {M.}~\bibnamefont {Poggio}},\ }\bibfield  {title} {\bibinfo {title}
  {Visualizing thickness-dependent magnetic textures in few-layer
  {Cr}$_{\textrm{2}}${Ge}$_{\textrm{2}}${Te}$_{\textrm{6}}$},\ }\href
  {https://doi.org/10.1038/s43246-024-00477-5} {\bibfield  {journal} {\bibinfo
  {journal} {Communications Materials}\ }\textbf {\bibinfo {volume} {5}},\
  \bibinfo {pages} {40} (\bibinfo {year} {2024})}\BibitemShut {NoStop}%
\bibitem [{\citenamefont {Bagani}\ \emph {et~al.}(2024)\citenamefont {Bagani},
  \citenamefont {Vervelaki}, \citenamefont {Jetter}, \citenamefont
  {Devarakonda}, \citenamefont {Tschudin}, \citenamefont {Gross}, \citenamefont
  {Chica}, \citenamefont {Broadway}, \citenamefont {Dean}, \citenamefont {Roy},
  \citenamefont {Maletinsky},\ and\ \citenamefont
  {Poggio}}]{bagani_imaging_2024}%
  \BibitemOpen
  \bibfield  {author} {\bibinfo {author} {\bibfnamefont {K.}~\bibnamefont
  {Bagani}}, \bibinfo {author} {\bibfnamefont {A.}~\bibnamefont {Vervelaki}},
  \bibinfo {author} {\bibfnamefont {D.}~\bibnamefont {Jetter}}, \bibinfo
  {author} {\bibfnamefont {A.}~\bibnamefont {Devarakonda}}, \bibinfo {author}
  {\bibfnamefont {M.~A.}\ \bibnamefont {Tschudin}}, \bibinfo {author}
  {\bibfnamefont {B.}~\bibnamefont {Gross}}, \bibinfo {author} {\bibfnamefont
  {D.~G.}\ \bibnamefont {Chica}}, \bibinfo {author} {\bibfnamefont {D.~A.}\
  \bibnamefont {Broadway}}, \bibinfo {author} {\bibfnamefont {C.~R.}\
  \bibnamefont {Dean}}, \bibinfo {author} {\bibfnamefont {X.}~\bibnamefont
  {Roy}}, \bibinfo {author} {\bibfnamefont {P.}~\bibnamefont {Maletinsky}},\
  and\ \bibinfo {author} {\bibfnamefont {M.}~\bibnamefont {Poggio}},\
  }\bibfield  {title} {\bibinfo {title} {Imaging strain-controlled magnetic
  reversal in thin {CrSBr}},\ }\href
  {https://doi.org/10.1021/acs.nanolett.4c03919} {\bibfield  {journal}
  {\bibinfo  {journal} {Nano Letters}\ }\textbf {\bibinfo {volume} {24}},\
  \bibinfo {pages} {13068} (\bibinfo {year} {2024})}\BibitemShut {NoStop}%
\bibitem [{\citenamefont {Tschirhart}\ \emph {et~al.}(2021)\citenamefont
  {Tschirhart}, \citenamefont {Serlin}, \citenamefont {Polshyn}, \citenamefont
  {Shragai}, \citenamefont {Xia}, \citenamefont {Zhu}, \citenamefont {Zhang},
  \citenamefont {Watanabe}, \citenamefont {Taniguchi}, \citenamefont {Huber},\
  and\ \citenamefont {Young}}]{tschirhart_imaging_2021}%
  \BibitemOpen
  \bibfield  {author} {\bibinfo {author} {\bibfnamefont {C.~L.}\ \bibnamefont
  {Tschirhart}}, \bibinfo {author} {\bibfnamefont {M.}~\bibnamefont {Serlin}},
  \bibinfo {author} {\bibfnamefont {H.}~\bibnamefont {Polshyn}}, \bibinfo
  {author} {\bibfnamefont {A.}~\bibnamefont {Shragai}}, \bibinfo {author}
  {\bibfnamefont {Z.}~\bibnamefont {Xia}}, \bibinfo {author} {\bibfnamefont
  {J.}~\bibnamefont {Zhu}}, \bibinfo {author} {\bibfnamefont {Y.}~\bibnamefont
  {Zhang}}, \bibinfo {author} {\bibfnamefont {K.}~\bibnamefont {Watanabe}},
  \bibinfo {author} {\bibfnamefont {T.}~\bibnamefont {Taniguchi}}, \bibinfo
  {author} {\bibfnamefont {M.~E.}\ \bibnamefont {Huber}},\ and\ \bibinfo
  {author} {\bibfnamefont {A.~F.}\ \bibnamefont {Young}},\ }\bibfield  {title}
  {\bibinfo {title} {Imaging orbital ferromagnetism in a moiré {Chern}
  insulator},\ }\href {https://doi.org/10.1126/science.abd3190} {\bibfield
  {journal} {\bibinfo  {journal} {Science}\ }\textbf {\bibinfo {volume}
  {372}},\ \bibinfo {pages} {1323} (\bibinfo {year} {2021})}\BibitemShut
  {NoStop}%
\bibitem [{\citenamefont {Nowack}\ \emph {et~al.}(2013)\citenamefont {Nowack},
  \citenamefont {Spanton}, \citenamefont {Baenninger}, \citenamefont {König},
  \citenamefont {Kirtley}, \citenamefont {Kalisky}, \citenamefont {Ames},
  \citenamefont {Leubner}, \citenamefont {Brüne}, \citenamefont {Buhmann},
  \citenamefont {Molenkamp}, \citenamefont {Goldhaber-Gordon},\ and\
  \citenamefont {Moler}}]{nowack_imaging_2013}%
  \BibitemOpen
  \bibfield  {author} {\bibinfo {author} {\bibfnamefont {K.~C.}\ \bibnamefont
  {Nowack}}, \bibinfo {author} {\bibfnamefont {E.~M.}\ \bibnamefont {Spanton}},
  \bibinfo {author} {\bibfnamefont {M.}~\bibnamefont {Baenninger}}, \bibinfo
  {author} {\bibfnamefont {M.}~\bibnamefont {König}}, \bibinfo {author}
  {\bibfnamefont {J.~R.}\ \bibnamefont {Kirtley}}, \bibinfo {author}
  {\bibfnamefont {B.}~\bibnamefont {Kalisky}}, \bibinfo {author} {\bibfnamefont
  {C.}~\bibnamefont {Ames}}, \bibinfo {author} {\bibfnamefont {P.}~\bibnamefont
  {Leubner}}, \bibinfo {author} {\bibfnamefont {C.}~\bibnamefont {Brüne}},
  \bibinfo {author} {\bibfnamefont {H.}~\bibnamefont {Buhmann}}, \bibinfo
  {author} {\bibfnamefont {L.~W.}\ \bibnamefont {Molenkamp}}, \bibinfo {author}
  {\bibfnamefont {D.}~\bibnamefont {Goldhaber-Gordon}},\ and\ \bibinfo {author}
  {\bibfnamefont {K.~A.}\ \bibnamefont {Moler}},\ }\bibfield  {title} {\bibinfo
  {title} {Imaging currents in {HgTe} quantum wells in the quantum spin {Hall}
  regime},\ }\href {https://doi.org/10.1038/nmat3682} {\bibfield  {journal}
  {\bibinfo  {journal} {Nature Materials}\ }\textbf {\bibinfo {volume} {12}},\
  \bibinfo {pages} {787} (\bibinfo {year} {2013})}\BibitemShut {NoStop}%
\bibitem [{\citenamefont {Ferguson}\ \emph {et~al.}(2023)\citenamefont
  {Ferguson}, \citenamefont {Xiao}, \citenamefont {Richardella}, \citenamefont
  {Low}, \citenamefont {Samarth},\ and\ \citenamefont
  {Nowack}}]{ferguson_direct_2023}%
  \BibitemOpen
  \bibfield  {author} {\bibinfo {author} {\bibfnamefont {G.~M.}\ \bibnamefont
  {Ferguson}}, \bibinfo {author} {\bibfnamefont {R.}~\bibnamefont {Xiao}},
  \bibinfo {author} {\bibfnamefont {A.~R.}\ \bibnamefont {Richardella}},
  \bibinfo {author} {\bibfnamefont {D.}~\bibnamefont {Low}}, \bibinfo {author}
  {\bibfnamefont {N.}~\bibnamefont {Samarth}},\ and\ \bibinfo {author}
  {\bibfnamefont {K.~C.}\ \bibnamefont {Nowack}},\ }\bibfield  {title}
  {\bibinfo {title} {Direct visualization of electronic transport in a quantum
  anomalous {Hall} insulator},\ }\href
  {https://doi.org/10.1038/s41563-023-01622-0} {\bibfield  {journal} {\bibinfo
  {journal} {Nature Materials}\ }\textbf {\bibinfo {volume} {22}},\ \bibinfo
  {pages} {1100} (\bibinfo {year} {2023})}\BibitemShut {NoStop}%
\bibitem [{\citenamefont {Uri}\ \emph {et~al.}(2020{\natexlab{a}})\citenamefont
  {Uri}, \citenamefont {Kim}, \citenamefont {Bagani}, \citenamefont
  {Lewandowski}, \citenamefont {Grover}, \citenamefont {Auerbach},
  \citenamefont {Lachman}, \citenamefont {Myasoedov}, \citenamefont
  {Taniguchi}, \citenamefont {Watanabe}, \citenamefont {Smet},\ and\
  \citenamefont {Zeldov}}]{uri_nanoscale_2020}%
  \BibitemOpen
  \bibfield  {author} {\bibinfo {author} {\bibfnamefont {A.}~\bibnamefont
  {Uri}}, \bibinfo {author} {\bibfnamefont {Y.}~\bibnamefont {Kim}}, \bibinfo
  {author} {\bibfnamefont {K.}~\bibnamefont {Bagani}}, \bibinfo {author}
  {\bibfnamefont {C.~K.}\ \bibnamefont {Lewandowski}}, \bibinfo {author}
  {\bibfnamefont {S.}~\bibnamefont {Grover}}, \bibinfo {author} {\bibfnamefont
  {N.}~\bibnamefont {Auerbach}}, \bibinfo {author} {\bibfnamefont {E.~O.}\
  \bibnamefont {Lachman}}, \bibinfo {author} {\bibfnamefont {Y.}~\bibnamefont
  {Myasoedov}}, \bibinfo {author} {\bibfnamefont {T.}~\bibnamefont
  {Taniguchi}}, \bibinfo {author} {\bibfnamefont {K.}~\bibnamefont {Watanabe}},
  \bibinfo {author} {\bibfnamefont {J.}~\bibnamefont {Smet}},\ and\ \bibinfo
  {author} {\bibfnamefont {E.}~\bibnamefont {Zeldov}},\ }\bibfield  {title}
  {\bibinfo {title} {Nanoscale imaging of equilibrium quantum {Hall} edge
  currents and of the magnetic monopole response in graphene},\ }\href
  {https://doi.org/10.1038/s41567-019-0713-3} {\bibfield  {journal} {\bibinfo
  {journal} {Nature Physics}\ }\textbf {\bibinfo {volume} {16}},\ \bibinfo
  {pages} {164} (\bibinfo {year} {2020}{\natexlab{a}})}\BibitemShut {NoStop}%
\bibitem [{\citenamefont {Uri}\ \emph {et~al.}(2020{\natexlab{b}})\citenamefont
  {Uri}, \citenamefont {Grover}, \citenamefont {Cao}, \citenamefont {Crosse},
  \citenamefont {Bagani}, \citenamefont {Rodan-Legrain}, \citenamefont
  {Myasoedov}, \citenamefont {Watanabe}, \citenamefont {Taniguchi},
  \citenamefont {Moon}, \citenamefont {Koshino}, \citenamefont
  {Jarillo-Herrero},\ and\ \citenamefont {Zeldov}}]{uri_mapping_2020}%
  \BibitemOpen
  \bibfield  {author} {\bibinfo {author} {\bibfnamefont {A.}~\bibnamefont
  {Uri}}, \bibinfo {author} {\bibfnamefont {S.}~\bibnamefont {Grover}},
  \bibinfo {author} {\bibfnamefont {Y.}~\bibnamefont {Cao}}, \bibinfo {author}
  {\bibfnamefont {J.~A.}\ \bibnamefont {Crosse}}, \bibinfo {author}
  {\bibfnamefont {K.}~\bibnamefont {Bagani}}, \bibinfo {author} {\bibfnamefont
  {D.}~\bibnamefont {Rodan-Legrain}}, \bibinfo {author} {\bibfnamefont
  {Y.}~\bibnamefont {Myasoedov}}, \bibinfo {author} {\bibfnamefont
  {K.}~\bibnamefont {Watanabe}}, \bibinfo {author} {\bibfnamefont
  {T.}~\bibnamefont {Taniguchi}}, \bibinfo {author} {\bibfnamefont
  {P.}~\bibnamefont {Moon}}, \bibinfo {author} {\bibfnamefont {M.}~\bibnamefont
  {Koshino}}, \bibinfo {author} {\bibfnamefont {P.}~\bibnamefont
  {Jarillo-Herrero}},\ and\ \bibinfo {author} {\bibfnamefont {E.}~\bibnamefont
  {Zeldov}},\ }\bibfield  {title} {\bibinfo {title} {Mapping the twist-angle
  disorder and {Landau} levels in magic-angle graphene},\ }\href
  {https://doi.org/10.1038/s41586-020-2255-3} {\bibfield  {journal} {\bibinfo
  {journal} {Nature}\ }\textbf {\bibinfo {volume} {581}},\ \bibinfo {pages}
  {47} (\bibinfo {year} {2020}{\natexlab{b}})}\BibitemShut {NoStop}%
\bibitem [{\citenamefont {Kalisky}\ \emph {et~al.}(2013)\citenamefont
  {Kalisky}, \citenamefont {Spanton}, \citenamefont {Noad}, \citenamefont
  {Kirtley}, \citenamefont {Nowack}, \citenamefont {Bell}, \citenamefont
  {Sato}, \citenamefont {Hosoda}, \citenamefont {Xie}, \citenamefont {Hikita},
  \citenamefont {Woltmann}, \citenamefont {Pfanzelt}, \citenamefont {Jany},
  \citenamefont {Richter}, \citenamefont {Hwang}, \citenamefont {Mannhart},\
  and\ \citenamefont {Moler}}]{kalisky_locally_2013}%
  \BibitemOpen
  \bibfield  {author} {\bibinfo {author} {\bibfnamefont {B.}~\bibnamefont
  {Kalisky}}, \bibinfo {author} {\bibfnamefont {E.~M.}\ \bibnamefont
  {Spanton}}, \bibinfo {author} {\bibfnamefont {H.}~\bibnamefont {Noad}},
  \bibinfo {author} {\bibfnamefont {J.~R.}\ \bibnamefont {Kirtley}}, \bibinfo
  {author} {\bibfnamefont {K.~C.}\ \bibnamefont {Nowack}}, \bibinfo {author}
  {\bibfnamefont {C.}~\bibnamefont {Bell}}, \bibinfo {author} {\bibfnamefont
  {H.~K.}\ \bibnamefont {Sato}}, \bibinfo {author} {\bibfnamefont
  {M.}~\bibnamefont {Hosoda}}, \bibinfo {author} {\bibfnamefont
  {Y.}~\bibnamefont {Xie}}, \bibinfo {author} {\bibfnamefont {Y.}~\bibnamefont
  {Hikita}}, \bibinfo {author} {\bibfnamefont {C.}~\bibnamefont {Woltmann}},
  \bibinfo {author} {\bibfnamefont {G.}~\bibnamefont {Pfanzelt}}, \bibinfo
  {author} {\bibfnamefont {R.}~\bibnamefont {Jany}}, \bibinfo {author}
  {\bibfnamefont {C.}~\bibnamefont {Richter}}, \bibinfo {author} {\bibfnamefont
  {H.~Y.}\ \bibnamefont {Hwang}}, \bibinfo {author} {\bibfnamefont
  {J.}~\bibnamefont {Mannhart}},\ and\ \bibinfo {author} {\bibfnamefont
  {K.~A.}\ \bibnamefont {Moler}},\ }\bibfield  {title} {\bibinfo {title}
  {Locally enhanced conductivity due to the tetragonal domain structure in
  {LaAlO}$_{\textrm{3}}$/{SrTiO}$_{\textrm{3}}$ heterointerfaces},\ }\href
  {https://doi.org/10.1038/nmat3753} {\bibfield  {journal} {\bibinfo  {journal}
  {Nature Materials}\ }\textbf {\bibinfo {volume} {12}},\ \bibinfo {pages}
  {1091} (\bibinfo {year} {2013})}\BibitemShut {NoStop}%
\bibitem [{\citenamefont {Aharon-Steinberg}\ \emph {et~al.}(2022)\citenamefont
  {Aharon-Steinberg}, \citenamefont {Völkl}, \citenamefont {Kaplan},
  \citenamefont {Pariari}, \citenamefont {Roy}, \citenamefont {Holder},
  \citenamefont {Wolf}, \citenamefont {Meltzer}, \citenamefont {Myasoedov},
  \citenamefont {Huber}, \citenamefont {Yan}, \citenamefont {Falkovich},
  \citenamefont {Levitov}, \citenamefont {Hücker},\ and\ \citenamefont
  {Zeldov}}]{aharon-steinberg_direct_2022}%
  \BibitemOpen
  \bibfield  {author} {\bibinfo {author} {\bibfnamefont {A.}~\bibnamefont
  {Aharon-Steinberg}}, \bibinfo {author} {\bibfnamefont {T.}~\bibnamefont
  {Völkl}}, \bibinfo {author} {\bibfnamefont {A.}~\bibnamefont {Kaplan}},
  \bibinfo {author} {\bibfnamefont {A.~K.}\ \bibnamefont {Pariari}}, \bibinfo
  {author} {\bibfnamefont {I.}~\bibnamefont {Roy}}, \bibinfo {author}
  {\bibfnamefont {T.}~\bibnamefont {Holder}}, \bibinfo {author} {\bibfnamefont
  {Y.}~\bibnamefont {Wolf}}, \bibinfo {author} {\bibfnamefont {A.~Y.}\
  \bibnamefont {Meltzer}}, \bibinfo {author} {\bibfnamefont {Y.}~\bibnamefont
  {Myasoedov}}, \bibinfo {author} {\bibfnamefont {M.~E.}\ \bibnamefont
  {Huber}}, \bibinfo {author} {\bibfnamefont {B.}~\bibnamefont {Yan}}, \bibinfo
  {author} {\bibfnamefont {G.}~\bibnamefont {Falkovich}}, \bibinfo {author}
  {\bibfnamefont {L.~S.}\ \bibnamefont {Levitov}}, \bibinfo {author}
  {\bibfnamefont {M.}~\bibnamefont {Hücker}},\ and\ \bibinfo {author}
  {\bibfnamefont {E.}~\bibnamefont {Zeldov}},\ }\bibfield  {title} {\bibinfo
  {title} {Direct observation of vortices in an electron fluid},\ }\href
  {https://doi.org/10.1038/s41586-022-04794-y} {\bibfield  {journal} {\bibinfo
  {journal} {Nature}\ }\textbf {\bibinfo {volume} {607}},\ \bibinfo {pages}
  {74} (\bibinfo {year} {2022})}\BibitemShut {NoStop}%
\bibitem [{\citenamefont {Persky}\ \emph
  {et~al.}(2022{\natexlab{a}})\citenamefont {Persky}, \citenamefont
  {Sochnikov},\ and\ \citenamefont {Kalisky}}]{persky_studying_2022}%
  \BibitemOpen
  \bibfield  {author} {\bibinfo {author} {\bibfnamefont {E.}~\bibnamefont
  {Persky}}, \bibinfo {author} {\bibfnamefont {I.}~\bibnamefont {Sochnikov}},\
  and\ \bibinfo {author} {\bibfnamefont {B.}~\bibnamefont {Kalisky}},\
  }\bibfield  {title} {\bibinfo {title} {Studying quantum materials with
  scanning {SQUID} microscopy},\ }\href
  {https://doi.org/10.1146/annurev-conmatphys-031620-104226} {\bibfield
  {journal} {\bibinfo  {journal} {Annual Review of Condensed Matter Physics}\
  }\textbf {\bibinfo {volume} {13}},\ \bibinfo {pages} {385} (\bibinfo {year}
  {2022}{\natexlab{a}})}\BibitemShut {NoStop}%
\bibitem [{\citenamefont {Kirtley}(2010)}]{kirtley_fundamental_2010}%
  \BibitemOpen
  \bibfield  {author} {\bibinfo {author} {\bibfnamefont {J.~R.}\ \bibnamefont
  {Kirtley}},\ }\bibfield  {title} {\bibinfo {title} {Fundamental studies of
  superconductors using scanning magnetic imaging},\ }\href
  {https://doi.org/10.1088/0034-4885/73/12/126501} {\bibfield  {journal}
  {\bibinfo  {journal} {Reports on Progress in Physics}\ }\textbf {\bibinfo
  {volume} {73}},\ \bibinfo {pages} {126501} (\bibinfo {year}
  {2010})}\BibitemShut {NoStop}%
\bibitem [{\citenamefont {Embon}\ \emph {et~al.}(2017)\citenamefont {Embon},
  \citenamefont {Anahory}, \citenamefont {Jelić}, \citenamefont {Lachman},
  \citenamefont {Myasoedov}, \citenamefont {Huber}, \citenamefont {Mikitik},
  \citenamefont {Silhanek}, \citenamefont {Milošević}, \citenamefont
  {Gurevich},\ and\ \citenamefont {Zeldov}}]{embon_imaging_2017}%
  \BibitemOpen
  \bibfield  {author} {\bibinfo {author} {\bibfnamefont {L.}~\bibnamefont
  {Embon}}, \bibinfo {author} {\bibfnamefont {Y.}~\bibnamefont {Anahory}},
  \bibinfo {author} {\bibfnamefont {{\v{Z}. L}.}~\bibnamefont {Jelić}},
  \bibinfo {author} {\bibfnamefont {E.~O.}\ \bibnamefont {Lachman}}, \bibinfo
  {author} {\bibfnamefont {Y.}~\bibnamefont {Myasoedov}}, \bibinfo {author}
  {\bibfnamefont {M.~E.}\ \bibnamefont {Huber}}, \bibinfo {author}
  {\bibfnamefont {G.~P.}\ \bibnamefont {Mikitik}}, \bibinfo {author}
  {\bibfnamefont {A.~V.}\ \bibnamefont {Silhanek}}, \bibinfo {author}
  {\bibfnamefont {M.~V.}\ \bibnamefont {Milošević}}, \bibinfo {author}
  {\bibfnamefont {A.}~\bibnamefont {Gurevich}},\ and\ \bibinfo {author}
  {\bibfnamefont {E.}~\bibnamefont {Zeldov}},\ }\bibfield  {title} {\bibinfo
  {title} {Imaging of super-fast dynamics and flow instabilities of
  superconducting vortices},\ }\href
  {https://doi.org/10.1038/s41467-017-00089-3} {\bibfield  {journal} {\bibinfo
  {journal} {Nature Communications}\ }\textbf {\bibinfo {volume} {8}},\
  \bibinfo {pages} {85} (\bibinfo {year} {2017})}\BibitemShut {NoStop}%
\bibitem [{\citenamefont {Kremen}\ \emph {et~al.}(2018)\citenamefont {Kremen},
  \citenamefont {Khan}, \citenamefont {Loh}, \citenamefont {Baturina},
  \citenamefont {Trivedi}, \citenamefont {Frydman},\ and\ \citenamefont
  {Kalisky}}]{kremen_imaging_2018}%
  \BibitemOpen
  \bibfield  {author} {\bibinfo {author} {\bibfnamefont {A.}~\bibnamefont
  {Kremen}}, \bibinfo {author} {\bibfnamefont {H.}~\bibnamefont {Khan}},
  \bibinfo {author} {\bibfnamefont {Y.~L.}\ \bibnamefont {Loh}}, \bibinfo
  {author} {\bibfnamefont {T.~I.}\ \bibnamefont {Baturina}}, \bibinfo {author}
  {\bibfnamefont {N.}~\bibnamefont {Trivedi}}, \bibinfo {author} {\bibfnamefont
  {A.}~\bibnamefont {Frydman}},\ and\ \bibinfo {author} {\bibfnamefont
  {B.}~\bibnamefont {Kalisky}},\ }\bibfield  {title} {\bibinfo {title} {Imaging
  quantum fluctuations near criticality},\ }\href
  {https://doi.org/10.1038/s41567-018-0264-z} {\bibfield  {journal} {\bibinfo
  {journal} {Nature Physics}\ }\textbf {\bibinfo {volume} {14}},\ \bibinfo
  {pages} {1205} (\bibinfo {year} {2018})}\BibitemShut {NoStop}%
\bibitem [{\citenamefont {Persky}\ \emph
  {et~al.}(2022{\natexlab{b}})\citenamefont {Persky}, \citenamefont {Bjørlig},
  \citenamefont {Feldman}, \citenamefont {Almoalem}, \citenamefont {Altman},
  \citenamefont {Berg}, \citenamefont {Kimchi}, \citenamefont {Ruhman},
  \citenamefont {Kanigel},\ and\ \citenamefont
  {Kalisky}}]{persky_magnetic_2022}%
  \BibitemOpen
  \bibfield  {author} {\bibinfo {author} {\bibfnamefont {E.}~\bibnamefont
  {Persky}}, \bibinfo {author} {\bibfnamefont {A.~V.}\ \bibnamefont
  {Bjørlig}}, \bibinfo {author} {\bibfnamefont {I.}~\bibnamefont {Feldman}},
  \bibinfo {author} {\bibfnamefont {A.}~\bibnamefont {Almoalem}}, \bibinfo
  {author} {\bibfnamefont {E.}~\bibnamefont {Altman}}, \bibinfo {author}
  {\bibfnamefont {E.}~\bibnamefont {Berg}}, \bibinfo {author} {\bibfnamefont
  {I.}~\bibnamefont {Kimchi}}, \bibinfo {author} {\bibfnamefont
  {J.}~\bibnamefont {Ruhman}}, \bibinfo {author} {\bibfnamefont
  {A.}~\bibnamefont {Kanigel}},\ and\ \bibinfo {author} {\bibfnamefont
  {B.}~\bibnamefont {Kalisky}},\ }\bibfield  {title} {\bibinfo {title}
  {Magnetic memory and spontaneous vortices in a van der {Waals}
  superconductor},\ }\href {https://doi.org/10.1038/s41586-022-04855-2}
  {\bibfield  {journal} {\bibinfo  {journal} {Nature}\ }\textbf {\bibinfo
  {volume} {607}},\ \bibinfo {pages} {692} (\bibinfo {year}
  {2022}{\natexlab{b}})}\BibitemShut {NoStop}%
\bibitem [{\citenamefont {Iguchi}\ \emph {et~al.}(2023)\citenamefont {Iguchi},
  \citenamefont {Shi}, \citenamefont {Kihou}, \citenamefont {Lee},
  \citenamefont {Barkman}, \citenamefont {Benfenati}, \citenamefont {Grinenko},
  \citenamefont {Babaev},\ and\ \citenamefont
  {Moler}}]{iguchi_superconducting_2023}%
  \BibitemOpen
  \bibfield  {author} {\bibinfo {author} {\bibfnamefont {Y.}~\bibnamefont
  {Iguchi}}, \bibinfo {author} {\bibfnamefont {R.~A.}\ \bibnamefont {Shi}},
  \bibinfo {author} {\bibfnamefont {K.}~\bibnamefont {Kihou}}, \bibinfo
  {author} {\bibfnamefont {C.-H.}\ \bibnamefont {Lee}}, \bibinfo {author}
  {\bibfnamefont {M.}~\bibnamefont {Barkman}}, \bibinfo {author} {\bibfnamefont
  {A.~L.}\ \bibnamefont {Benfenati}}, \bibinfo {author} {\bibfnamefont
  {V.}~\bibnamefont {Grinenko}}, \bibinfo {author} {\bibfnamefont
  {E.}~\bibnamefont {Babaev}},\ and\ \bibinfo {author} {\bibfnamefont {K.~A.}\
  \bibnamefont {Moler}},\ }\bibfield  {title} {\bibinfo {title}
  {Superconducting vortices carrying a temperature-dependent fraction of the
  flux quantum},\ }\href {https://doi.org/10.1126/science.abp9979} {\bibfield
  {journal} {\bibinfo  {journal} {Science}\ }\textbf {\bibinfo {volume}
  {380}},\ \bibinfo {pages} {1244} (\bibinfo {year} {2023})}\BibitemShut
  {NoStop}%
\bibitem [{\citenamefont {Kirtley}\ \emph {et~al.}(1995)\citenamefont
  {Kirtley}, \citenamefont {Ketchen}, \citenamefont {Stawiasz}, \citenamefont
  {Sun}, \citenamefont {Gallagher}, \citenamefont {Blanton},\ and\
  \citenamefont {Wind}}]{kirtley_highresolution_1995}%
  \BibitemOpen
  \bibfield  {author} {\bibinfo {author} {\bibfnamefont {J.~R.}\ \bibnamefont
  {Kirtley}}, \bibinfo {author} {\bibfnamefont {M.~B.}\ \bibnamefont
  {Ketchen}}, \bibinfo {author} {\bibfnamefont {K.~G.}\ \bibnamefont
  {Stawiasz}}, \bibinfo {author} {\bibfnamefont {J.~Z.}\ \bibnamefont {Sun}},
  \bibinfo {author} {\bibfnamefont {W.~J.}\ \bibnamefont {Gallagher}}, \bibinfo
  {author} {\bibfnamefont {S.~H.}\ \bibnamefont {Blanton}},\ and\ \bibinfo
  {author} {\bibfnamefont {S.~J.}\ \bibnamefont {Wind}},\ }\bibfield  {title}
  {\bibinfo {title} {High‐resolution scanning {SQUID} microscope},\ }\href
  {https://doi.org/10.1063/1.113838} {\bibfield  {journal} {\bibinfo  {journal}
  {Applied Physics Letters}\ }\textbf {\bibinfo {volume} {66}},\ \bibinfo
  {pages} {1138} (\bibinfo {year} {1995})}\BibitemShut {NoStop}%
\bibitem [{\citenamefont {Koshnick}\ \emph {et~al.}(2008)\citenamefont
  {Koshnick}, \citenamefont {Huber}, \citenamefont {Bert}, \citenamefont
  {Hicks}, \citenamefont {Large}, \citenamefont {Edwards},\ and\ \citenamefont
  {Moler}}]{koshnick_terraced_2008}%
  \BibitemOpen
  \bibfield  {author} {\bibinfo {author} {\bibfnamefont {N.~C.}\ \bibnamefont
  {Koshnick}}, \bibinfo {author} {\bibfnamefont {M.~E.}\ \bibnamefont {Huber}},
  \bibinfo {author} {\bibfnamefont {J.~A.}\ \bibnamefont {Bert}}, \bibinfo
  {author} {\bibfnamefont {C.~W.}\ \bibnamefont {Hicks}}, \bibinfo {author}
  {\bibfnamefont {J.}~\bibnamefont {Large}}, \bibinfo {author} {\bibfnamefont
  {H.}~\bibnamefont {Edwards}},\ and\ \bibinfo {author} {\bibfnamefont {K.~A.}\
  \bibnamefont {Moler}},\ }\bibfield  {title} {\bibinfo {title} {A terraced
  scanning super conducting quantum interference device susceptometer with
  submicron pickup loops},\ }\href {https://doi.org/10.1063/1.3046098}
  {\bibfield  {journal} {\bibinfo  {journal} {Applied Physics Letters}\
  }\textbf {\bibinfo {volume} {93}},\ \bibinfo {pages} {243101} (\bibinfo
  {year} {2008})}\BibitemShut {NoStop}%
\bibitem [{\citenamefont {Kirtley}\ \emph {et~al.}(2016)\citenamefont
  {Kirtley}, \citenamefont {Paulius}, \citenamefont {Rosenberg}, \citenamefont
  {Palmstrom}, \citenamefont {Holland}, \citenamefont {Spanton}, \citenamefont
  {Schiessl}, \citenamefont {Jermain}, \citenamefont {Gibbons}, \citenamefont
  {Fung}, \citenamefont {Huber}, \citenamefont {Ralph}, \citenamefont
  {Ketchen}, \citenamefont {Gibson},\ and\ \citenamefont
  {Moler}}]{kirtley_scanning_2016}%
  \BibitemOpen
  \bibfield  {author} {\bibinfo {author} {\bibfnamefont {J.~R.}\ \bibnamefont
  {Kirtley}}, \bibinfo {author} {\bibfnamefont {L.}~\bibnamefont {Paulius}},
  \bibinfo {author} {\bibfnamefont {A.~J.}\ \bibnamefont {Rosenberg}}, \bibinfo
  {author} {\bibfnamefont {J.~C.}\ \bibnamefont {Palmstrom}}, \bibinfo {author}
  {\bibfnamefont {C.~M.}\ \bibnamefont {Holland}}, \bibinfo {author}
  {\bibfnamefont {E.~M.}\ \bibnamefont {Spanton}}, \bibinfo {author}
  {\bibfnamefont {D.}~\bibnamefont {Schiessl}}, \bibinfo {author}
  {\bibfnamefont {C.~L.}\ \bibnamefont {Jermain}}, \bibinfo {author}
  {\bibfnamefont {J.}~\bibnamefont {Gibbons}}, \bibinfo {author} {\bibfnamefont
  {Y.-K.-K.}\ \bibnamefont {Fung}}, \bibinfo {author} {\bibfnamefont {M.~E.}\
  \bibnamefont {Huber}}, \bibinfo {author} {\bibfnamefont {D.~C.}\ \bibnamefont
  {Ralph}}, \bibinfo {author} {\bibfnamefont {M.~B.}\ \bibnamefont {Ketchen}},
  \bibinfo {author} {\bibfnamefont {G.~W.}\ \bibnamefont {Gibson}},\ and\
  \bibinfo {author} {\bibfnamefont {K.~A.}\ \bibnamefont {Moler}},\ }\bibfield
  {title} {\bibinfo {title} {Scanning {SQUID} susceptometers with sub-micron
  spatial resolution},\ }\href {https://doi.org/10.1063/1.4961982} {\bibfield
  {journal} {\bibinfo  {journal} {Review of Scientific Instruments}\ }\textbf
  {\bibinfo {volume} {87}},\ \bibinfo {pages} {093702} (\bibinfo {year}
  {2016})}\BibitemShut {NoStop}%
\bibitem [{\citenamefont {Huber}\ \emph {et~al.}(2008)\citenamefont {Huber},
  \citenamefont {Koshnick}, \citenamefont {Bluhm}, \citenamefont {Archuleta},
  \citenamefont {Azua}, \citenamefont {Björnsson}, \citenamefont {Gardner},
  \citenamefont {Halloran}, \citenamefont {Lucero},\ and\ \citenamefont
  {Moler}}]{huber_gradiometric_2008}%
  \BibitemOpen
  \bibfield  {author} {\bibinfo {author} {\bibfnamefont {M.~E.}\ \bibnamefont
  {Huber}}, \bibinfo {author} {\bibfnamefont {N.~C.}\ \bibnamefont {Koshnick}},
  \bibinfo {author} {\bibfnamefont {H.}~\bibnamefont {Bluhm}}, \bibinfo
  {author} {\bibfnamefont {L.~J.}\ \bibnamefont {Archuleta}}, \bibinfo {author}
  {\bibfnamefont {T.}~\bibnamefont {Azua}}, \bibinfo {author} {\bibfnamefont
  {P.~G.}\ \bibnamefont {Björnsson}}, \bibinfo {author} {\bibfnamefont
  {B.~W.}\ \bibnamefont {Gardner}}, \bibinfo {author} {\bibfnamefont {S.~T.}\
  \bibnamefont {Halloran}}, \bibinfo {author} {\bibfnamefont {E.~A.}\
  \bibnamefont {Lucero}},\ and\ \bibinfo {author} {\bibfnamefont {K.~A.}\
  \bibnamefont {Moler}},\ }\bibfield  {title} {\bibinfo {title} {Gradiometric
  micro-{SQUID} susceptometer for scanning measurements of mesoscopic
  samples},\ }\href {https://doi.org/10.1063/1.2932341} {\bibfield  {journal}
  {\bibinfo  {journal} {Review of Scientific Instruments}\ }\textbf {\bibinfo
  {volume} {79}},\ \bibinfo {pages} {053704} (\bibinfo {year}
  {2008})}\BibitemShut {NoStop}%
\bibitem [{\citenamefont {Finkler}\ \emph {et~al.}(2010)\citenamefont
  {Finkler}, \citenamefont {Segev}, \citenamefont {Myasoedov}, \citenamefont
  {Rappaport}, \citenamefont {Ne’eman}, \citenamefont {Vasyukov},
  \citenamefont {Zeldov}, \citenamefont {Huber}, \citenamefont {Martin},\ and\
  \citenamefont {Yacoby}}]{finkler_self-aligned_2010}%
  \BibitemOpen
  \bibfield  {author} {\bibinfo {author} {\bibfnamefont {A.}~\bibnamefont
  {Finkler}}, \bibinfo {author} {\bibfnamefont {Y.}~\bibnamefont {Segev}},
  \bibinfo {author} {\bibfnamefont {Y.}~\bibnamefont {Myasoedov}}, \bibinfo
  {author} {\bibfnamefont {M.~L.}\ \bibnamefont {Rappaport}}, \bibinfo {author}
  {\bibfnamefont {L.}~\bibnamefont {Ne’eman}}, \bibinfo {author}
  {\bibfnamefont {D.}~\bibnamefont {Vasyukov}}, \bibinfo {author}
  {\bibfnamefont {E.}~\bibnamefont {Zeldov}}, \bibinfo {author} {\bibfnamefont
  {M.~E.}\ \bibnamefont {Huber}}, \bibinfo {author} {\bibfnamefont
  {J.}~\bibnamefont {Martin}},\ and\ \bibinfo {author} {\bibfnamefont
  {A.}~\bibnamefont {Yacoby}},\ }\bibfield  {title} {\bibinfo {title}
  {Self-aligned nanoscale {SQUID} on a tip},\ }\href
  {https://doi.org/10.1021/nl100009r} {\bibfield  {journal} {\bibinfo
  {journal} {Nano Letters}\ }\textbf {\bibinfo {volume} {10}},\ \bibinfo
  {pages} {1046} (\bibinfo {year} {2010})}\BibitemShut {NoStop}%
\bibitem [{\citenamefont {Vasyukov}\ \emph {et~al.}(2013)\citenamefont
  {Vasyukov}, \citenamefont {Anahory}, \citenamefont {Embon}, \citenamefont
  {Halbertal}, \citenamefont {Cuppens}, \citenamefont {Neeman}, \citenamefont
  {Finkler}, \citenamefont {Segev}, \citenamefont {Myasoedov}, \citenamefont
  {Rappaport}, \citenamefont {Huber},\ and\ \citenamefont
  {Zeldov}}]{vasyukov_scanning_2013}%
  \BibitemOpen
  \bibfield  {author} {\bibinfo {author} {\bibfnamefont {D.}~\bibnamefont
  {Vasyukov}}, \bibinfo {author} {\bibfnamefont {Y.}~\bibnamefont {Anahory}},
  \bibinfo {author} {\bibfnamefont {L.}~\bibnamefont {Embon}}, \bibinfo
  {author} {\bibfnamefont {D.}~\bibnamefont {Halbertal}}, \bibinfo {author}
  {\bibfnamefont {J.}~\bibnamefont {Cuppens}}, \bibinfo {author} {\bibfnamefont
  {L.}~\bibnamefont {Neeman}}, \bibinfo {author} {\bibfnamefont
  {A.}~\bibnamefont {Finkler}}, \bibinfo {author} {\bibfnamefont
  {Y.}~\bibnamefont {Segev}}, \bibinfo {author} {\bibfnamefont
  {Y.}~\bibnamefont {Myasoedov}}, \bibinfo {author} {\bibfnamefont {M.~L.}\
  \bibnamefont {Rappaport}}, \bibinfo {author} {\bibfnamefont {M.~E.}\
  \bibnamefont {Huber}},\ and\ \bibinfo {author} {\bibfnamefont
  {E.}~\bibnamefont {Zeldov}},\ }\bibfield  {title} {\bibinfo {title} {A
  scanning superconducting quantum interference device with single electron
  spin sensitivity},\ }\href {https://doi.org/10.1038/nnano.2013.169}
  {\bibfield  {journal} {\bibinfo  {journal} {Nature Nanotechnology}\ }\textbf
  {\bibinfo {volume} {8}},\ \bibinfo {pages} {639} (\bibinfo {year}
  {2013})}\BibitemShut {NoStop}%
\bibitem [{\citenamefont {Bagani}\ \emph {et~al.}(2019)\citenamefont {Bagani},
  \citenamefont {Sarkar}, \citenamefont {Uri}, \citenamefont {Rappaport},
  \citenamefont {Huber}, \citenamefont {Zeldov},\ and\ \citenamefont
  {Myasoedov}}]{bagani_sputtered_2019}%
  \BibitemOpen
  \bibfield  {author} {\bibinfo {author} {\bibfnamefont {K.}~\bibnamefont
  {Bagani}}, \bibinfo {author} {\bibfnamefont {J.}~\bibnamefont {Sarkar}},
  \bibinfo {author} {\bibfnamefont {A.}~\bibnamefont {Uri}}, \bibinfo {author}
  {\bibfnamefont {M.~L.}\ \bibnamefont {Rappaport}}, \bibinfo {author}
  {\bibfnamefont {M.~E.}\ \bibnamefont {Huber}}, \bibinfo {author}
  {\bibfnamefont {E.}~\bibnamefont {Zeldov}},\ and\ \bibinfo {author}
  {\bibfnamefont {Y.}~\bibnamefont {Myasoedov}},\ }\bibfield  {title} {\bibinfo
  {title} {Sputtered {Mo}$_{\textrm{66}}${Re}$_{\textrm{34}}$ {SQUID}-on-{Tip}
  for high-field magnetic and thermal nanoimaging},\ }\href
  {https://doi.org/10.1103/PhysRevApplied.12.044062} {\bibfield  {journal}
  {\bibinfo  {journal} {Physical Review Applied}\ }\textbf {\bibinfo {volume}
  {12}},\ \bibinfo {pages} {044062} (\bibinfo {year} {2019})}\BibitemShut
  {NoStop}%
\bibitem [{\citenamefont {José Martínez-Pérez}\ and\ \citenamefont
  {Koelle}(2017)}]{jose_martinez-perez_nanosquids_2017}%
  \BibitemOpen
  \bibfield  {author} {\bibinfo {author} {\bibfnamefont {M.}~\bibnamefont
  {José Martínez-Pérez}}\ and\ \bibinfo {author} {\bibfnamefont
  {D.}~\bibnamefont {Koelle}},\ }\bibfield  {title} {\bibinfo {title}
  {{NanoSQUIDs}: {Basics} \& recent advances},\ }\href
  {https://doi.org/10.1515/psr-2017-5001} {\bibfield  {journal} {\bibinfo
  {journal} {Physical Sciences Reviews}\ }\textbf {\bibinfo {volume} {2}},\
  \bibinfo {pages} {20175001} (\bibinfo {year} {2017})}\BibitemShut {NoStop}%
\bibitem [{\citenamefont {Wyss}\ \emph {et~al.}(2022)\citenamefont {Wyss},
  \citenamefont {Bagani}, \citenamefont {Jetter}, \citenamefont {Marchiori},
  \citenamefont {Vervelaki}, \citenamefont {Gross}, \citenamefont {Ridderbos},
  \citenamefont {Gliga}, \citenamefont {Schönenberger},\ and\ \citenamefont
  {Poggio}}]{wyss_magnetic_2022}%
  \BibitemOpen
  \bibfield  {author} {\bibinfo {author} {\bibfnamefont {M.}~\bibnamefont
  {Wyss}}, \bibinfo {author} {\bibfnamefont {K.}~\bibnamefont {Bagani}},
  \bibinfo {author} {\bibfnamefont {D.}~\bibnamefont {Jetter}}, \bibinfo
  {author} {\bibfnamefont {E.}~\bibnamefont {Marchiori}}, \bibinfo {author}
  {\bibfnamefont {A.}~\bibnamefont {Vervelaki}}, \bibinfo {author}
  {\bibfnamefont {B.}~\bibnamefont {Gross}}, \bibinfo {author} {\bibfnamefont
  {J.}~\bibnamefont {Ridderbos}}, \bibinfo {author} {\bibfnamefont
  {S.}~\bibnamefont {Gliga}}, \bibinfo {author} {\bibfnamefont
  {C.}~\bibnamefont {Schönenberger}},\ and\ \bibinfo {author} {\bibfnamefont
  {M.}~\bibnamefont {Poggio}},\ }\bibfield  {title} {\bibinfo {title}
  {Magnetic, thermal, and topographic imaging with a nanometer-scale
  {SQUID}-on-lever scanning probe},\ }\href
  {https://doi.org/10.1103/PhysRevApplied.17.034002} {\bibfield  {journal}
  {\bibinfo  {journal} {Physical Review Applied}\ }\textbf {\bibinfo {volume}
  {17}},\ \bibinfo {pages} {034002} (\bibinfo {year} {2022})}\BibitemShut
  {NoStop}%
\bibitem [{\citenamefont {Weber}\ \emph {et~al.}(2025)\citenamefont {Weber},
  \citenamefont {Cadorim}, \citenamefont {Milošević}, \citenamefont
  {Kleiner},\ and\ \citenamefont {Koelle}}]{weber_niobium_2025}%
  \BibitemOpen
  \bibfield  {author} {\bibinfo {author} {\bibfnamefont {T.}~\bibnamefont
  {Weber}}, \bibinfo {author} {\bibfnamefont {L.}~\bibnamefont {Cadorim}},
  \bibinfo {author} {\bibfnamefont {M.~V.}\ \bibnamefont {Milošević}},
  \bibinfo {author} {\bibfnamefont {R.}~\bibnamefont {Kleiner}},\ and\ \bibinfo
  {author} {\bibfnamefont {D.}~\bibnamefont {Koelle}},\ }\bibfield  {title}
  {\bibinfo {title} {Niobium {Dayem} bridges fabricated by {Ne} and {He}
  focused ion beams},\ }\href@noop {} {\bibfield  {journal} {\bibinfo
  {journal} {Unpublished}\ } (\bibinfo {year} {2025})}\BibitemShut {NoStop}%
\bibitem [{\citenamefont {Meltzer}\ \emph {et~al.}(2016)\citenamefont
  {Meltzer}, \citenamefont {Uri},\ and\ \citenamefont
  {Zeldov}}]{meltzer_multi-terminal_2016}%
  \BibitemOpen
  \bibfield  {author} {\bibinfo {author} {\bibfnamefont {A.~Y.}\ \bibnamefont
  {Meltzer}}, \bibinfo {author} {\bibfnamefont {A.}~\bibnamefont {Uri}},\ and\
  \bibinfo {author} {\bibfnamefont {E.}~\bibnamefont {Zeldov}},\ }\bibfield
  {title} {\bibinfo {title} {Multi-terminal multi-junction dc {SQUID} for
  nanoscale magnetometry},\ }\href
  {https://doi.org/10.1088/0953-2048/29/11/114001} {\bibfield  {journal}
  {\bibinfo  {journal} {Superconductor Science and Technology}\ }\textbf
  {\bibinfo {volume} {29}},\ \bibinfo {pages} {114001} (\bibinfo {year}
  {2016})}\BibitemShut {NoStop}%
\bibitem [{\citenamefont {Uri}\ \emph {et~al.}(2016)\citenamefont {Uri},
  \citenamefont {Meltzer}, \citenamefont {Anahory}, \citenamefont {Embon},
  \citenamefont {Lachman}, \citenamefont {Halbertal}, \citenamefont {HR},
  \citenamefont {Myasoedov}, \citenamefont {Huber}, \citenamefont {Young},\
  and\ \citenamefont {Zeldov}}]{uri_electrically_2016}%
  \BibitemOpen
  \bibfield  {author} {\bibinfo {author} {\bibfnamefont {A.}~\bibnamefont
  {Uri}}, \bibinfo {author} {\bibfnamefont {A.~Y.}\ \bibnamefont {Meltzer}},
  \bibinfo {author} {\bibfnamefont {Y.}~\bibnamefont {Anahory}}, \bibinfo
  {author} {\bibfnamefont {L.}~\bibnamefont {Embon}}, \bibinfo {author}
  {\bibfnamefont {E.~O.}\ \bibnamefont {Lachman}}, \bibinfo {author}
  {\bibfnamefont {D.}~\bibnamefont {Halbertal}}, \bibinfo {author}
  {\bibfnamefont {N.}~\bibnamefont {HR}}, \bibinfo {author} {\bibfnamefont
  {Y.}~\bibnamefont {Myasoedov}}, \bibinfo {author} {\bibfnamefont {M.~E.}\
  \bibnamefont {Huber}}, \bibinfo {author} {\bibfnamefont {A.~F.}\ \bibnamefont
  {Young}},\ and\ \bibinfo {author} {\bibfnamefont {E.}~\bibnamefont
  {Zeldov}},\ }\bibfield  {title} {\bibinfo {title} {Electrically tunable
  multiterminal {SQUID}-on-{Tip}},\ }\href
  {https://doi.org/10.1021/acs.nanolett.6b02841} {\bibfield  {journal}
  {\bibinfo  {journal} {Nano Letters}\ }\textbf {\bibinfo {volume} {16}},\
  \bibinfo {pages} {6910} (\bibinfo {year} {2016})}\BibitemShut {NoStop}%
\bibitem [{\citenamefont {Wolter}\ \emph {et~al.}(2022)\citenamefont {Wolter},
  \citenamefont {Linek}, \citenamefont {Weimann}, \citenamefont {Koelle},
  \citenamefont {Kleiner},\ and\ \citenamefont {Kieler}}]{wolter_static_2022}%
  \BibitemOpen
  \bibfield  {author} {\bibinfo {author} {\bibfnamefont {S.}~\bibnamefont
  {Wolter}}, \bibinfo {author} {\bibfnamefont {J.}~\bibnamefont {Linek}},
  \bibinfo {author} {\bibfnamefont {T.}~\bibnamefont {Weimann}}, \bibinfo
  {author} {\bibfnamefont {D.}~\bibnamefont {Koelle}}, \bibinfo {author}
  {\bibfnamefont {R.}~\bibnamefont {Kleiner}},\ and\ \bibinfo {author}
  {\bibfnamefont {O.}~\bibnamefont {Kieler}},\ }\bibfield  {title} {\bibinfo
  {title} {Static and dynamic transport properties of multi-terminal,
  multi-junction {microSQUIDs} realized with {Nb}/{HfTi}/{Nb} {Josephson}
  junctions},\ }\href {https://doi.org/10.1088/1361-6668/ac782b} {\bibfield
  {journal} {\bibinfo  {journal} {Superconductor Science and Technology}\
  }\textbf {\bibinfo {volume} {35}},\ \bibinfo {pages} {085006} (\bibinfo
  {year} {2022})}\BibitemShut {NoStop}%
\bibitem [{\citenamefont {Tesche}\ and\ \citenamefont
  {Clarke}(1977)}]{tesche_dc_1977}%
  \BibitemOpen
  \bibfield  {author} {\bibinfo {author} {\bibfnamefont {C.~D.}\ \bibnamefont
  {Tesche}}\ and\ \bibinfo {author} {\bibfnamefont {J.}~\bibnamefont
  {Clarke}},\ }\bibfield  {title} {\bibinfo {title} {dc {SQUID}: {Noise} and
  optimization},\ }\href {https://doi.org/10.1007/BF00655097} {\bibfield
  {journal} {\bibinfo  {journal} {Journal of Low Temperature Physics}\ }\textbf
  {\bibinfo {volume} {29}},\ \bibinfo {pages} {301} (\bibinfo {year}
  {1977})}\BibitemShut {NoStop}%
\bibitem [{\citenamefont {Chesca}\ \emph {et~al.}(2004)\citenamefont {Chesca},
  \citenamefont {Kleiner},\ and\ \citenamefont {Koelle}}]{chesca_SQUID_2004}%
  \BibitemOpen
  \bibfield  {author} {\bibinfo {author} {\bibfnamefont {B.}~\bibnamefont
  {Chesca}}, \bibinfo {author} {\bibfnamefont {R.}~\bibnamefont {Kleiner}},\
  and\ \bibinfo {author} {\bibfnamefont {D.}~\bibnamefont {Koelle}},\ }\bibinfo
  {title} {{SQUID} {Theory}},\ in\ \href
  {https://doi.org/10.1002/3527603646.ch2} {\emph {\bibinfo {booktitle} {The
  {SQUID} {Handbook}}}}\ (\bibinfo  {publisher} {John Wiley \& Sons, Ltd},\
  \bibinfo {year} {2004})\ pp.\ \bibinfo {pages} {29--92}\BibitemShut {NoStop}%
\bibitem [{\citenamefont {Prance}\ and\ \citenamefont
  {Thompson}(2023)}]{prance_sensitivity_2023}%
  \BibitemOpen
  \bibfield  {author} {\bibinfo {author} {\bibfnamefont {J.~R.}\ \bibnamefont
  {Prance}}\ and\ \bibinfo {author} {\bibfnamefont {M.~D.}\ \bibnamefont
  {Thompson}},\ }\bibfield  {title} {\bibinfo {title} {Sensitivity of a {DC}
  {SQUID} with a non-sinusoidal current-phase relation in its junctions},\
  }\href {https://doi.org/10.1063/5.0151607} {\bibfield  {journal} {\bibinfo
  {journal} {Applied Physics Letters}\ }\textbf {\bibinfo {volume} {122}},\
  \bibinfo {pages} {222601} (\bibinfo {year} {2023})}\BibitemShut {NoStop}%
\bibitem [{\citenamefont {Khapaev}\ \emph {et~al.}(2002)\citenamefont
  {Khapaev}, \citenamefont {Kupriyanov}, \citenamefont {Goldobin},\ and\
  \citenamefont {Siegel}}]{khapaev_current_2002}%
  \BibitemOpen
  \bibfield  {author} {\bibinfo {author} {\bibfnamefont {M.~M.}\ \bibnamefont
  {Khapaev}}, \bibinfo {author} {\bibfnamefont {M.~Y.}\ \bibnamefont
  {Kupriyanov}}, \bibinfo {author} {\bibfnamefont {E.}~\bibnamefont
  {Goldobin}},\ and\ \bibinfo {author} {\bibfnamefont {M.}~\bibnamefont
  {Siegel}},\ }\bibfield  {title} {\bibinfo {title} {Current distribution
  simulation for superconducting multi-layered structures},\ }\href
  {https://doi.org/10.1088/0953-2048/16/1/305} {\bibfield  {journal} {\bibinfo
  {journal} {Superconductor Science and Technology}\ }\textbf {\bibinfo
  {volume} {16}},\ \bibinfo {pages} {24} (\bibinfo {year} {2002})}\BibitemShut
  {NoStop}%
\bibitem [{\citenamefont {Marchiori}\ \emph {et~al.}(2024)\citenamefont
  {Marchiori}, \citenamefont {Romagnoli}, \citenamefont {Schneider},
  \citenamefont {Gross}, \citenamefont {Sahafi}, \citenamefont {Jordan},
  \citenamefont {Budakian}, \citenamefont {Baral}, \citenamefont {Magrez},
  \citenamefont {White},\ and\ \citenamefont
  {Poggio}}]{marchiori_imaging_2024}%
  \BibitemOpen
  \bibfield  {author} {\bibinfo {author} {\bibfnamefont {E.}~\bibnamefont
  {Marchiori}}, \bibinfo {author} {\bibfnamefont {G.}~\bibnamefont
  {Romagnoli}}, \bibinfo {author} {\bibfnamefont {L.}~\bibnamefont
  {Schneider}}, \bibinfo {author} {\bibfnamefont {B.}~\bibnamefont {Gross}},
  \bibinfo {author} {\bibfnamefont {P.}~\bibnamefont {Sahafi}}, \bibinfo
  {author} {\bibfnamefont {A.}~\bibnamefont {Jordan}}, \bibinfo {author}
  {\bibfnamefont {R.}~\bibnamefont {Budakian}}, \bibinfo {author}
  {\bibfnamefont {P.~R.}\ \bibnamefont {Baral}}, \bibinfo {author}
  {\bibfnamefont {A.}~\bibnamefont {Magrez}}, \bibinfo {author} {\bibfnamefont
  {J.~S.}\ \bibnamefont {White}},\ and\ \bibinfo {author} {\bibfnamefont
  {M.}~\bibnamefont {Poggio}},\ }\bibfield  {title} {\bibinfo {title} {Imaging
  magnetic spiral phases, skyrmion clusters, and skyrmion displacements at the
  surface of bulk {Cu}$_{\textrm{2}}${OSeO}$_{\textrm{3}}$},\ }\href
  {https://doi.org/10.1038/s43246-024-00647-5} {\bibfield  {journal} {\bibinfo
  {journal} {Communications Materials}\ }\textbf {\bibinfo {volume} {5}},\
  \bibinfo {pages} {202} (\bibinfo {year} {2024})}\BibitemShut {NoStop}%
\bibitem [{\citenamefont {Marchiori}\ \emph {et~al.}(2022)\citenamefont
  {Marchiori}, \citenamefont {Ceccarelli}, \citenamefont {Rossi}, \citenamefont
  {Lorenzelli}, \citenamefont {Degen},\ and\ \citenamefont
  {Poggio}}]{marchiori_nanoscale_2022}%
  \BibitemOpen
  \bibfield  {author} {\bibinfo {author} {\bibfnamefont {E.}~\bibnamefont
  {Marchiori}}, \bibinfo {author} {\bibfnamefont {L.}~\bibnamefont
  {Ceccarelli}}, \bibinfo {author} {\bibfnamefont {N.}~\bibnamefont {Rossi}},
  \bibinfo {author} {\bibfnamefont {L.}~\bibnamefont {Lorenzelli}}, \bibinfo
  {author} {\bibfnamefont {C.~L.}\ \bibnamefont {Degen}},\ and\ \bibinfo
  {author} {\bibfnamefont {M.}~\bibnamefont {Poggio}},\ }\bibfield  {title}
  {\bibinfo {title} {Nanoscale magnetic field imaging for {2D} materials},\
  }\href {https://doi.org/10.1038/s42254-021-00380-9} {\bibfield  {journal}
  {\bibinfo  {journal} {Nature Reviews Physics}\ }\textbf {\bibinfo {volume}
  {4}},\ \bibinfo {pages} {49} (\bibinfo {year} {2022})}\BibitemShut {NoStop}%
\bibitem [{\citenamefont {Vansteenkiste}\ \emph {et~al.}(2014)\citenamefont
  {Vansteenkiste}, \citenamefont {Leliaert}, \citenamefont {Dvornik},
  \citenamefont {Helsen}, \citenamefont {Garcia-Sanchez},\ and\ \citenamefont
  {Van~Waeyenberge}}]{vansteenkiste_design_2014}%
  \BibitemOpen
  \bibfield  {author} {\bibinfo {author} {\bibfnamefont {A.}~\bibnamefont
  {Vansteenkiste}}, \bibinfo {author} {\bibfnamefont {J.}~\bibnamefont
  {Leliaert}}, \bibinfo {author} {\bibfnamefont {M.}~\bibnamefont {Dvornik}},
  \bibinfo {author} {\bibfnamefont {M.}~\bibnamefont {Helsen}}, \bibinfo
  {author} {\bibfnamefont {F.}~\bibnamefont {Garcia-Sanchez}},\ and\ \bibinfo
  {author} {\bibfnamefont {B.}~\bibnamefont {Van~Waeyenberge}},\ }\bibfield
  {title} {\bibinfo {title} {The design and verification of {MuMax3}},\ }\href
  {https://doi.org/10.1063/1.4899186} {\bibfield  {journal} {\bibinfo
  {journal} {AIP Advances}\ }\textbf {\bibinfo {volume} {4}},\ \bibinfo {pages}
  {107133} (\bibinfo {year} {2014})}\BibitemShut {NoStop}%
\bibitem [{\citenamefont {Exl}\ \emph {et~al.}(2014)\citenamefont {Exl},
  \citenamefont {Bance}, \citenamefont {Reichel}, \citenamefont {Schrefl},
  \citenamefont {Peter~Stimming},\ and\ \citenamefont
  {Mauser}}]{exl_labontes_2014}%
  \BibitemOpen
  \bibfield  {author} {\bibinfo {author} {\bibfnamefont {L.}~\bibnamefont
  {Exl}}, \bibinfo {author} {\bibfnamefont {S.}~\bibnamefont {Bance}}, \bibinfo
  {author} {\bibfnamefont {F.}~\bibnamefont {Reichel}}, \bibinfo {author}
  {\bibfnamefont {T.}~\bibnamefont {Schrefl}}, \bibinfo {author} {\bibfnamefont
  {H.}~\bibnamefont {Peter~Stimming}},\ and\ \bibinfo {author} {\bibfnamefont
  {N.~J.}\ \bibnamefont {Mauser}},\ }\bibfield  {title} {\bibinfo {title}
  {{LaBonte}'s method revisited: {An} effective steepest descent method for
  micromagnetic energy minimization},\ }\href
  {https://doi.org/10.1063/1.4862839} {\bibfield  {journal} {\bibinfo
  {journal} {Journal of Applied Physics}\ }\textbf {\bibinfo {volume} {115}},\
  \bibinfo {pages} {17D118} (\bibinfo {year} {2014})}\BibitemShut {NoStop}%
\bibitem [{\citenamefont {Combettes}\ and\ \citenamefont
  {Pesquet}(2011)}]{combettes_proximal_2011}%
  \BibitemOpen
  \bibfield  {author} {\bibinfo {author} {\bibfnamefont {P.~L.}\ \bibnamefont
  {Combettes}}\ and\ \bibinfo {author} {\bibfnamefont {J.-C.}\ \bibnamefont
  {Pesquet}},\ }\bibinfo {title} {Proximal splitting methods in signal
  processing},\ in\ \href {https://doi.org/10.1007/978-1-4419-9569-8_10} {\emph
  {\bibinfo {booktitle} {Fixed-Point Algorithms for Inverse Problems in Science
  and Engineering}}},\ \bibinfo {editor} {edited by\ \bibinfo {editor}
  {\bibfnamefont {H.~H.}\ \bibnamefont {Bauschke}}, \bibinfo {editor}
  {\bibfnamefont {R.~S.}\ \bibnamefont {Burachik}}, \bibinfo {editor}
  {\bibfnamefont {P.~L.}\ \bibnamefont {Combettes}}, \bibinfo {editor}
  {\bibfnamefont {V.}~\bibnamefont {Elser}}, \bibinfo {editor} {\bibfnamefont
  {D.~R.}\ \bibnamefont {Luke}},\ and\ \bibinfo {editor} {\bibfnamefont
  {H.}~\bibnamefont {Wolkowicz}}}\ (\bibinfo  {publisher} {Springer New York},\
  \bibinfo {address} {New York, NY},\ \bibinfo {year} {2011})\ pp.\ \bibinfo
  {pages} {185--212}\BibitemShut {NoStop}%
\bibitem [{\citenamefont {Adams}\ \emph {et~al.}(2012)\citenamefont {Adams},
  \citenamefont {Chacon}, \citenamefont {Wagner}, \citenamefont {Bauer},
  \citenamefont {Brandl}, \citenamefont {Pedersen}, \citenamefont {Berger},
  \citenamefont {Lemmens},\ and\ \citenamefont
  {Pfleiderer}}]{adams_long-wavelength_2012}%
  \BibitemOpen
  \bibfield  {author} {\bibinfo {author} {\bibfnamefont {T.}~\bibnamefont
  {Adams}}, \bibinfo {author} {\bibfnamefont {A.}~\bibnamefont {Chacon}},
  \bibinfo {author} {\bibfnamefont {M.}~\bibnamefont {Wagner}}, \bibinfo
  {author} {\bibfnamefont {A.}~\bibnamefont {Bauer}}, \bibinfo {author}
  {\bibfnamefont {G.}~\bibnamefont {Brandl}}, \bibinfo {author} {\bibfnamefont
  {B.}~\bibnamefont {Pedersen}}, \bibinfo {author} {\bibfnamefont
  {H.}~\bibnamefont {Berger}}, \bibinfo {author} {\bibfnamefont
  {P.}~\bibnamefont {Lemmens}},\ and\ \bibinfo {author} {\bibfnamefont
  {C.}~\bibnamefont {Pfleiderer}},\ }\bibfield  {title} {\bibinfo {title}
  {Long-wavelength helimagnetic order and skyrmion lattice phase in
  {Cu}$_{\textrm{2}}${OSeO}$_{\textrm{3}}$},\ }\href
  {https://doi.org/10.1103/PhysRevLett.108.237204} {\bibfield  {journal}
  {\bibinfo  {journal} {Physical Review Letters}\ }\textbf {\bibinfo {volume}
  {108}},\ \bibinfo {pages} {237204} (\bibinfo {year} {2012})}\BibitemShut
  {NoStop}%
\bibitem [{\citenamefont {Baral}\ \emph {et~al.}(2022)\citenamefont {Baral},
  \citenamefont {Ukleev}, \citenamefont {LaGrange}, \citenamefont {Cubitt},
  \citenamefont {Živković}, \citenamefont {Rønnow}, \citenamefont {White},\
  and\ \citenamefont {Magrez}}]{baral_tuning_2022}%
  \BibitemOpen
  \bibfield  {author} {\bibinfo {author} {\bibfnamefont {P.~R.}\ \bibnamefont
  {Baral}}, \bibinfo {author} {\bibfnamefont {V.}~\bibnamefont {Ukleev}},
  \bibinfo {author} {\bibfnamefont {T.}~\bibnamefont {LaGrange}}, \bibinfo
  {author} {\bibfnamefont {R.}~\bibnamefont {Cubitt}}, \bibinfo {author}
  {\bibfnamefont {I.}~\bibnamefont {Živković}}, \bibinfo {author}
  {\bibfnamefont {H.~M.}\ \bibnamefont {Rønnow}}, \bibinfo {author}
  {\bibfnamefont {J.~S.}\ \bibnamefont {White}},\ and\ \bibinfo {author}
  {\bibfnamefont {A.}~\bibnamefont {Magrez}},\ }\bibfield  {title} {\bibinfo
  {title} {Tuning topological spin textures in size-tailored chiral magnet
  insulator particles},\ }\href {https://doi.org/10.1021/acs.jpcc.2c03600}
  {\bibfield  {journal} {\bibinfo  {journal} {The Journal of Physical Chemistry
  C}\ }\textbf {\bibinfo {volume} {126}},\ \bibinfo {pages} {11855} (\bibinfo
  {year} {2022})}\BibitemShut {NoStop}%
\end{thebibliography}%

\end{document}